\documentclass{aastex}
\usepackage{emulateapj5}

\shorttitle{Iron and $\alpha$-abundances in Carina}
\shortauthors{Koch et al.}

\slugcomment{To appear in the Astronomical Journal}

\begin{document}

\title{Complexity on Small Scales III: \\ Iron and $\alpha$ Element Abundances 
in the Carina Dwarf Spheroidal Galaxy\altaffilmark{1}}

\author{Andreas Koch\altaffilmark{2,3}, Eva K.~Grebel\altaffilmark{2,4}, 
	Gerard F.~Gilmore\altaffilmark{5}, Rosemary F.G.~Wyse\altaffilmark{6}, \\
        Jan T.~Kleyna\altaffilmark{7}, 
	Daniel R.~Harbeck\altaffilmark{8},  
	Mark I.~Wilkinson\altaffilmark{9}, 
	and N.~Wyn Evans\altaffilmark{5}}
\email{akoch@astro.ucla.edu}

\altaffiltext{1}{Based on observations collected at the European Southern 
Observatory at Paranal, Chile; Large Programme proposal 171.B-0520(A).}
\altaffiltext{2}{Department of Physics and Astronomy, Astronomical Institute of the University of Basel,
Binningen, Switzerland}
\altaffiltext{3}{UCLA, Department of Physics and Astronomy, Los Angeles, CA, USA}
\altaffiltext{4}{Astronomisches Rechen-Institut, Zentrum f\"ur Astronomie Heidelberg,
University of Heidelberg,  Heidelberg, Germany}
\altaffiltext{5}{Institute of Astronomy, Cambridge University, Cambridge, UK}
\altaffiltext{6}{The John Hopkins University, Baltimore, MD, USA}
\altaffiltext{7}{Institute for Astronomy, University of Hawaii, Honolulu, HI, USA}
\altaffiltext{8}{Department of Astronomy, University of Wisconsin, Madison, WI, USA}
\altaffiltext{9}{Department of Physics and Astronomy, University of Leicester, Leicester, UK}

\begin{abstract}
  We have obtained high-resolution spectroscopy of ten red giants in
  the Carina dwarf spheroidal (dSph) galaxy with the UVES spectrograph
  at the ESO Very Large Telescope in order to study the detailed
  chemical evolution of this Galactic satellite. Here we present the
  abundances of O, Na, Mg, Si, Ca, Ti and Fe.  By comparison of
  the derived iron abundances [Fe/H] with metallicities based on the
  well established calcium triplet (CaT) calibration, [Fe/H]$_{\rm
    CaT}$, we show that the empirical CaT technique yields good
  agreement with the high-resolution data for [Fe/H]\,$\ga-2$ dex, but
  tends to deviate from these data at lower metallicities.  With
  [Fe/H]$\sim-1.7$ dex the mean iron abundance of our targets is fully
  consistent with the peak metallicity of Carina as derived from
  medium-resolution spectroscopy and previous photometric studies, all
  calibrated onto iron via Galactic globular cluster scales. We
  identify two metal poor stars with iron abundances of $-2.72$ and
  $-2.50$ dex.  These stars are found to have enhanced [$\alpha$/Fe]
  ratios similar to the elemental ratios of stars in the Milky Way
  halo. In this context it is conceivable that the moderately metal
  poor halo stars may originate from an early dSph accretion event.
  The bulk of the Carina red giants exhibit a depletion in
  the [$\alpha$/Fe] abundance ratios with respect to the Galactic halo
  at a given metallicity.  One of our targets with a moderately low
  [Fe/H] of $-1.5$ dex is considerably depleted in almost all of the
  $\alpha$-elements by $\sim0.5$ dex compared to the solar
  values. Such low values of the ratio of $\alpha$-elements to iron
  can be produced by stochastical fluctuations in terms of an
  incomplete mixing of single Type Ia and Type II supernova (SN)
  events into the interstellar medium. Moreover, the system's slow
  star formation (SF) rate grants sufficient time for SNe\,I to occur.
  Our derived chemical element ratios are consistent with the episodic
  and extended SF in Carina previously derived from
  analyses of its color-magnitude diagram. We find a considerable
  star-to-star scatter in the abundance ratios.  This suggests 
  that Carina's SF history varies with position within the
  galaxy, with incomplete mixing. In addition, or alternatively, the SF rate
  is so low that the high-mass stellar IMF is sparsely populated, as 
  expected on statistical grounds in low-mass star clusters, 
  leading to true scatter in the resultant mass-integrated
  yields. Both ideas are consistent with slow stochastic SF in
  dissolving associations or star clusters, so that one may not speak
  {\em prima facie} of a single ``SF history'' at a detailed
  level.
\end{abstract}

\keywords{Galaxies: abundances --- Galaxies: dwarf --- Galaxies: evolution 
--- Galaxies: stellar  content --- Galaxies: structure --- 
Galaxies: individual (\objectname{Carina}) --- Local Group}

\section{Introduction}
Dwarf spheroidal galaxies (dSphs) are intriguing objects: they are the
least luminous, least massive galaxies known (e.g., Grebel et al.
2003).  They also appear to be the most dark-matter dominated objects
(e.g., Gilmore et al.\ 2007 and references therein).  For many years
dSphs were considered the potential building blocks of larger
galaxies, and thus potential key stones from which to extract the
merging history of the Milky Way (e.g., Gallagher \& Wyse 1994 and
references therein).  While there is clear evidence for past mergers
of dwarf galaxies with the Milky Way, the age distributions and
metallicity distributions of stars in the dSphs, plus
recent detailed abundance analyses (Grebel \& Gallagher 2004; Venn et
al 2004), have called a simple building block scenario in question. 
These limit accretion and merging of systems like the surviving
dSphs to early epochs only (Unavane et al. 1996).

Among the dSphs orbiting the Milky Way the Carina dSph stands out
because of its unusual star formation (SF) history. It is the only dSph to
exhibit clearly episodic SF interrupted by long
quiescent periods.  Carina has been the subject of numerous, primarily
photometric, studies (for an overview of past work, see Koch et al.\
2006a; hereafter Paper I).  We recently analysed Carina's metallicity
distribution function (MDF) derived from medium-resolution
spectroscopy around the near-infrared Calcium II triplet (CaT) of 437
red giants (Paper I), from which we found a mean metallicity of
[Fe/H]$_{\rm CaT} = -1.7$ dex on the scale of Carretta \& Gratton (1997),
with a wide range from about $-3.0$ to 0.0 dex. Hence, there is evidence of
considerable enrichment during the evolutionary life time of this dSph
from about 12\,Gyr ago to a few Gyr ago.

Low- to medium resolution spectroscopic studies are well suited to
investigate the true, spectroscopic mean metallicity and metallicity
spread of individual dSphs and to derive their 
MDF. Combined with photometry, this permits
one to assess to what extent age and metallicity conspire (i.e., the
age-metallicity degeneracy in color-space), to search for possible
spatial gradients, and to explore the galactic evolutionary histories
taking into account their chemical evolution. One should note that the
overall `metallicity', which is commonly represented via [Fe/H] or
[Fe/H]$_{\rm CaT}$, is generally most sensitive to the {\em
  integrated} past SF history (see, e.g., Paper\,I), while it is
rather insensitive to the details of that SF history. Unravelling SF
histories requires knowledge of individual elemental
abundances and of the associated stellar ages: 
the evolution of specific chemical element ratios is
intimately linked to the SF history of a galaxy, the initial mass
function (IMF), and the binary fraction governing the rate of Supernovae (SNe) of Type Ia.

An important example of such abundance ratios is [$\alpha$/Fe], which
can be efficiently used as a ``chemical clock'' (Tinsley 1976; Wyse \&
Gilmore 1988; Matteucci 2003).  The $\alpha$-elements (O, Mg, Si, Ca,
Ti) form in quickly evolving high-mass stars, which end their lives in
core-collapse SNe II. Fe is also synthesized
in SNe\,II, but predominantly in SN\,Ia, which occur on much longer
time scales.  Hence, modulo loss of metals from the galaxy, the
[$\alpha$/Fe]-ratio of a a stellar population records its SF 
 history, reflecting the relative delay in the injection of
Fe and $\alpha$-elements into the interstellar medium (Matteucci \&
Greggio 1986; Gilmore \& Wyse 1998; Lanfranchi \& Matteucci 2004;
Lanfranchi et al.\ 2006; Robertson et al.\ 2005; Font et al.\ 2006).

Stars in dSphs have lower [$\alpha$/Fe] abundance ratios on average
than do Galactic halo stars of the same metallicity (Shetrone,
C\^ot\'e \& Sargent 2001; Fulbright 2002, Shetrone et al.\ 2003
[hereafter S03]; Tolstoy et al.\ 2003 [hereafter T03]; Geisler et al.\
2005).  This is generally interpreted as evidence of a low SF rate in
dSphs compared to the Galactic halo, as predicted by Unavane et
al. (1996). The actual values will sensitively depend on the SF
and ISM mixing histories, and the IMFs, of both the dSph galaxy and
of the formation location of the field halo stars.

From their analysis of five red giants in Carina, S03 and T03 deduce
the imprints of SF bursts and pauses in Carina's abundance patterns.
Each SF burst can be expected to be accompanied by a rapid increase in
[$\alpha$/Fe] (Gilmore \& Wyse 1991, 1998), followed by a smoother
decline in this ratio during any subsequent period of low SF.  
This simply reflects the continuing production of Fe in
SNe\,Ia, which continue for many Gyr after the formation of their
progenitor stars (Matteucci \& Greggio 1986), with little
$\alpha$-enrichment due to the paucity of SNe\,II.  This SF 
and consequent enrichment history was claimed to be also
reflected in variations of the heavy-element patterns, such as some
[$s$/$r$]-process element abundance ratios vs.\ metallicity.
The illustrative scenarios proposed by S03 and T03 suggest an ancient
epoch of SF (around 13\,Gyr ago), a subsequent quiescent phase of some
3\,Gyr, and another SF period approximately 7--11\,Gyr ago.  This is,
however, not fully compatible with Carina's detailed SF history as
derived from color-magnitude diagram (CMD) modeling (e.g.,
Smecker-Hane et al.\ 1994; Mighell et al.\ 1997; Hurley-Keller et al.\
1998; Monelli et al.\ 2003; Rizzi et al.\ 2003) and thus merits
further investigation.

Our current study is a further step into this direction.  We have
enlarged the previously available sample by adding ten new stars, for
which we obtained high-resolution spectra using the UVES
spectrograph at the Very Large Telescope (VLT) of the European
Southern Observatory.  Considering the considerable expense in exposure
time for high-resolution studies, this represents already a
substantial improvement, even though these data still only yield
snapshots in sampling the SF episodes of Carina.  Each of our target
stars provides a valuable picture of its local environment at the time
of its formation. We have underway a much larger VLT/UVES study of
Carina, to extend this work.
 
Calibration of the reduced width of the near-infrared CaT index in
terms of metallicity is another issue involving an understanding of
the abundance of the $\alpha$-elements.  The widely
used, empirical calibrations (Armandroff \& Da Costa 1991; Rutledge et
al.\ 1997a,b) are all based on the metallicity scales of Galactic
globular clusters (e.g., Zinn \& West 1984; Carretta \& Gratton 1997;
Kraft \& Ivans 2003), themselves calibrated by iron elemental
abundances derived from high-resolution echelle spectra.  Cole et al.\
(2004) have extended this calibration to ages as young as 2.5 Gyr and
metallicities in the range of $-2.0 < $\,[Fe/H]\,$ < -0.2$ dex so that 
even the metallicities of stars in dSphs with extended SF histories
and prominent intermediate-age populations can be measured from the
CaT.  However, [$\alpha$/Fe] varies systematically across this range
in the calibrators, so that it remains an open question whether and to
what extent the extrapolation of the calibrations towards the
metal-poor and metal-rich tails of the MDFs 
introduces a possible bias or significant deviations from the ``true''
Fe-line stellar metallicity.  The CaT calibration explicitly assumes
a [Ca/Fe] ratio that is similar to the one found in the Galactic
globular clusters, i.e., a ratio of the order $\sim+0.3$ dex (Carney
1996; McWilliam 1997; Pritzl et al.\ 2005).  However, in dSphs this
ratio is not an {\em a priori} known quantity, and generally this
value shows a wide range, even within one galaxy.  There is evidence
that some stars in some dSphs can reach [Ca/Fe] values as enhanced as
the Galactic halo (Shetrone et al.\ 2001, S03).  Bosler et
al. (2007) argue that the CaT line strength is naturally best mapped
onto [Ca/H], where the respective calibration is independent of the SF
history of the system and valid over a broad range of metallicities
and ages.
Our measurements of [$\alpha$/Fe] in Carina will then aid in
elucidating the validity and accuracy of traditional CaT metallicity
measurements in the dSphs.

In this current study we focus on the abundance patterns for Na, Fe
and the $\alpha$-elements O, Mg, Si, Ca, Ti in ten red giants in
Carina.  An analysis of the heavy element abundance ratios will be
presented in a forthcoming paper (Koch et al., in preparation).
In \textsection 2, we describe our data and their reduction.
\textsection 3 describes the atomic data and atmospheric parameters
used in the abundance analysis including a detailed discussion of
errors.  In \textsection 4 we present our abundance results for iron
and the individual $\alpha$-elements before discussing their
implications for Carina's chemical evolution in \textsection 5.
\textsection 6 summarizes our findings. 

\section{Data and reduction}

Our data were obtained in the course of the ESO Large Programme
171.B-0520(A) (PI: G.~F.\ Gilmore; see also Paper I; Wyse et al.\
2006; Wilkinson et al.\ 2006; Koch et al. 2007, for details) that aims
at studying the kinematic and chemical characteristics of Galactic
dSphs. Two of these galaxies were extensively studied in medium
resolution mode, namely Leo\,II (Koch et al.\ 2007) and Carina
(Paper\,I).  In parallel, a number of high-resolution spectra of red
giants in Carina were obtained, in order to study the detailed
chemical composition of this particular galaxy. To allow us to perform
accurate chemical abundance studies at a high signal-to-noise (S/N) ratio
within a reasonable integration time, 30 bright giants below the tip
of the RGB were selected from the CMD, reaching V-band magnitudes
slightly above 18\,mag (see Table~1). Eleven of these stars were also
observed with medium-resolution around the CaT region, from which we
derived Carina's MDF in Paper\,I. For these stars
we can compare directly the stellar metallicities from the CaT
calibration with the iron abundance from the high-resolution
measurements, and thus assess the validity of the CaT calibration
method.

\subsection{Data acquisition}

In parallel with the observing runs in December 2003 (see Paper\,I),
during which medium-resolution spectroscopy of red giants in Carina
was obtained, we used the high-resolution spectrograph UVES
(Ultraviolet and Visual Echelle Spectrograph) at the ESO/VLT in
multi-object mode.  Using UVES in combination with the FLAMES fibre
facility permits one to feed nine fibres to the UVES instrument, where
we dedicated one fibre per setup to observing blank sky. This
configuration facilitates the sky subtraction later on.

The FLAMES/UVES fibre mode has been designed for the red arm of the
spectrograph. Our observations used the standard setup with a central
wavelength of 580\,nm, which gives a total spectral range of
470--680\,nm.  With a fibre aperture on the sky of 1$\arcsec$ and our
chosen CCD binning of 1$\times$1 pixels the instrument provides a
resolving power of $\sim$47000.

\subsection{Data reduction}

The data were reduced using the standard UVES pipeline (Modigliani
2004\footnote{This document is available online via\\
\url{http://www.eso.org/projects/dfs/dfs-shared/web/vlt/vlt-instrument-pipelines.html}.}),
which operates within the MIDAS environment\footnote{The European
Southern Observatory Munich Image Data Analysis System (ESO-MIDAS) is
developed and maintained by the European Southern Observatory.}.  The
respective standard steps comprise bias subtraction and division by a
master slit flat field.  The definition of the spectral orders is
obtained by the usual means of processing odd- and even-numbered fibre
flat field frames, combined with a first guess order position table
from the Th-Ar wavelength calibration lamp.  In the next step, the
wavelength solution is determined via calibration exposures from this
Th-Ar lamp, followed by the extraction of the target (and sky) spectra
using an optimal extraction algorithm.  The final extracted,
wavelength-calibrated spectra with contiguously merged orders are then
sky-subtracted using the dedicated sky spectrum taken in each setup.
Finally, the individual exposures are shifted to the heliocentric
standard of rest and coadded, weighted by the spectrum's individual
S/N.  Typical S/N ratios of the final spectra achieved in this way are
approximately 20\,pixel$^{-1}$.

\subsection{Membership estimates}

The likely Carina membership of the target stars was assessed by means
of their radial velocities.  We generated a template spectrum from a
line list comprising 101 strong absorption features with equivalent
widths (EWs) as expected for typical red giant spectra in dSphs.
Cross-correlation of the observations against this template using {\sc
iraf}'s\footnote{{\sc iraf} is distributed by the National Optical
Astronomy Observatories, which are operated by the Association of
Universities for Research in Astronomy, Inc., under cooperative
agreement with the National Science Foundation.} {\em fxcor} task
finally yielded accurate radial velocities with a median uncertainty
of 0.9\,km\,s$^{-1}$. 

For the 11 targets for which we also obtained medium-resolution
spectra of the CaT region, the velocity measurements from the UVES
spectra are in excellent agreement with the velocities derived
from the CaT (see Paper\,I) to within their uncertainties, and we find
a median offset between both estimates of 0.5\,km\,s$^{-1}$ with a 1$\sigma$-scatter 
of 1.1\,km\,s$^{-1}$.  

Out of our 30 targets, 12 stars clearly peak at Carina's systemic
velocity around a mean of 225\,km\,s$^{-1}$ with a dispersion of
6.4\,km\,s$^{-1}$. This is in good agreement with Carina's radial
velocity distribution (Mateo 1998; Majewski et al.\ 2005; Paper\,I).
The remaining 18 targets have radial velocities that deviate by more
than 120\,km\,s$^{-1}$ from this mean, which corresponds to
20\,$\sigma$ and unambiguously rules out their membership in Carina.
These stars are not considered in the current study and
are likely Galactic interlopers (Wyse et al.\ 2006). We note that the
candidates for observation were selected deliberately to extend well
to the blue and to the red of the Carina RGB, so that any very
metal-poor or metal-rich Carina stars would be included. No such stars
were found.
Two of the radial velocity members were identified as carbon stars via
their strong molecular bands.  These spectra were omitted from the  
abundance analyses presented here. 

\section{Abundance analysis}

For the abundance determinations, we employed the common technique of
an EW analysis. For this purpose, we used the 2002
version of the stellar abundance code MOOG (Sneden 1973), in
particular using its {\em abfind} driver.  For the van der Waals line
damping, the standard Uns\"old approximation was adopted within MOOG.

The accurate derivation of stellar abundance ratios sensitively relies
on the choice of proper atomic data, in particular the oscillator
strengths.  Concern about using solar $gf$-values for the analysis of
our late-type red giant target stars prompted us to perform a
differential abundance analysis relative to Arcturus ($\alpha$\,Boo),
a bright K1.5 disk giant of only moderate metal-deficiency
([Fe/H]=$-0.50$ dex) and with atmospheric parameters very similar to
those expected for the Carina red giants of our sample (e.g., Peterson
et al.\ 1993; Friel et al.\ 2003; Fulbright et al.\ 2006, 2007; Koch \&
McWilliam 2008).  For this purpose, we used the Arcturus spectral
atlas of Hinkle et al. (2000) to obtain a set of reference
measurements.

\subsection{Equivalent widths and $gf$-values}

In order to identify suitable transitions for our abundance
determinations, we compiled a line list from various literature
sources. Essentially, we exploited the compilations from the bulge
giant and thick disk star studies of McWilliam et al.\ (1995) and
Prochaska et al.\ (2000), from which we assembled a number of
absorption features for a number of chemical elements (see references
in these works for the sources of the respective atomic data).  We
chose the lines with the least amount of stellar and telluric
contamination, which thus appear reasonably unblended.

The EWs of these lines were measured using {\sc IRAF}'s {\em
splot}-task assuming a single Gaussian line profile. 
Typical {\em splot}-errors on the widths are of the order of
5\% (a few m\AA), and the EWs of those lines in common 
with  Fulbright et al. (2006, 2007) agree to within 3\% 
with their measurements. 

The oscillator strengths used in the majority of abundance studies are
nowadays based on laboratory measurements, such as the O'Brian values
(O'Brian et al.\ 1991), Hannover measurements (Bard \& Kock 1994) or
the Oxford $gf$-values (Blackwell et al.\ 1995), all of which are 
employed in the work of McWilliam et al.\ (1995) and
Prochaska et al.\ (2000). Still, as Prochaska et al.\ (2000) note,
even with such maximally accurate laboratory measurements at hand, the
$gf$-values impose a major error source in red giant abundance
studies, particularly when referencing one's abundances to the solar
meteoritic (Anders \& Grevesse 1989) or the updated empirical
(Grevesse \& Sauval 1999; Asplund et al.\ 2005) solar abundance
reference frame.  Therefore we have chosen to conduct our analysis
differentially with respect to the well-studied red giant Arcturus,
thus minimizing the uncertainties associated with solar and laboratory
$gf$-values.  Since Arcturus exhibits atmospheric
properties similar to the stars of our own study, such an analysis
will also aid the reduction of systematic errors, such as a priori unknown
deficiencies in the model atmospheres or the influence of weak,
unknown blends, which might scale with metallicity (e.g., Fulbright et
al.\ 2006).  To this end we measured the EWs of the lines in Arcturus from
the high-quality spectral atlas of Hinkle et al.\ (2000).

Adopting the well established atmospheric parameters of $\alpha$\,Boo
(T$_{\mathrm{eff}}$\,=\,4290\,K, log\,$g$\,=\,1.55,
$\xi$\,=\,1.67\,km\,s$^{-1}$, [Fe/H]\,=\,$-$0.5) from Fulbright et
al.\ (2006) and our measured EWs, we derived the $gf$-values on a
line-by-line basis in order to reproduce each element's abundance from
Fulbright et al.\ (2006, 2007).  As a result, our Arcturus-based
oscillator strengths agree with the laboratory values to within
$\sim$8\% on average.  The full line list is given in Table~2,
together with the respective EW measurements for each star.
   
\subsection{Model Atmospheres and stellar parameters}

Throughout our analyses we interpolated the model atmospheres from the
updated grid of R.~L. Kurucz's\footnote{These atmosphere grids can be
  downloaded from \url{http://cfaku5.cfa.harvard.edu/grids.html}}
one-dimensional 72-layer plane-parallel line-blanketed models with the
convective overshoot option switched off and assuming that local
thermodynamic equilibrium (LTE) holds for all species.  Moreover, our
model calculations incorporated the new opacity distribution
functions, {\sc odfnew}, provided by F.\ Castelli\footnote{See
  \url{http://wwwuser.oat.ts.astro.it/castelli}.}  (Castelli \& Kurucz
2003) for the Carina stars, assuming scaled-solar elemental
abundances, and the respective $\alpha$-element enhanced models, {\sc
  aodfnew}, for Arcturus. The latter are necessary since it is known
that Arcturus exhibits a considerable enhancement in these elements
(Peterson et al.\ 1993; Fulbright et al.\ 2007) of
[$\alpha$/Fe]$\sim+0.4\,$dex.

The initial estimates for the stellar effective temperatures were derived using dereddened
colors from the photometry obtained by the ESO Imaging Survey
(EIS\footnote{The reduced and calibrated photometry is available from
the EIS Web pages, see \url{http://www.eso.org/science/eis/}.}; Nonino
et al.\ 1999; Paper\,I), complemented by infrared colors, which were
drawn from the Point Source Catalogue of 2MASS (Cutri 2003). 
For Carina, we use a reddening of E\,(B$-$V)\,=\,0.06 (Mighell 1997),
which is also consistent with the value we found from the extinction
maps of Schlegel et al.\ (1998). To correct the infrared colors, the
respective reddening relations from Cardelli et al.\ (1989) were
applied. Finally, photometric effective temperatures were calibrated
using the empirical relations derived by Alonso et al.\ (1999, 2001).
Since the Alonso et al.\ calibrations of B$-$V and V$-$K require
colors in the Johnson-Cousins (JC) system, we employed the photometric
transformations described in Paper\,I in order to convert the input B
and V magnitudes from the photometric system of the EIS to the
standard JC system.  The 2MASS K-band magnitude 
was transformed to the Bessel \& Brett (1988) homogenized system,
using the equations provided by the Explanatory Supplement to the
2MASS All Sky Data Release (Cutri 2003).  Since this transformation
includes two steps of calibrations, i.e., from the 2MASS to the CIT
system and from CIT to the Bessel \& Brett (1998) system, the
transformation errors from both calibrations have to be incorporated
to yield the final photometric uncertainty of the colors (see also
Sivarani et al.\ 2004).  Table~3 lists the photometric properties of
all target stars with radial velocities that are consistent with
membership in Carina.

As initial guesses for the metallicity to be used in the
T$_{\mathrm{eff}}$\ calibrations and in the model atmospheres we
adopted the value derived from our CaT study of Paper\,I for the six
member stars in common.  Otherwise Carina's mean metallicity of $-1.7$
dex (Paper\,I) was assumed for the remaining four red giants for which
no metallicity measurement was available.

In practice, we adopted the T$_{\mathrm{eff}}$\,(V$-$K) value as the
final stellar photometric temperature, since the large wavelength
baseline of this color index makes this temperature the least
susceptible to the underlying photometric uncertainties, yielding the
most accurate results. The temperatures obtained from the B$-$V colors
are systematically cooler than their respective V$-$K counterparts by
$\sim$70\,K on average.  The random errors on the 
effective temperature, based on the full photometric uncertainties,
amount to 70\,K on average.
 
The stellar surface gravities, log\,$g$, were initially determined
using the basic stellar structure equations 
\begin{equation}
  \log\frac{g}{g_{\odot}}\,=\,\log\frac{M}{M_{\odot}}\,
+\,4\log\frac{{\rm
      T}_{\rm eff}}{{\rm T}_{{\rm eff},\,\odot}}\,-\,0.4\,(M_{{\rm
      bol},\,\odot}\,-\,M_V\,-\,{\rm BC}), 
\end{equation} 
where we adopted T$_{{\rm eff},\,\odot}=5777$\,K, $M_{{\rm
    bol},\,\odot}=4.74$ and log\,$g_{\odot}=4.44$ and a distance of
Carina of 94\,kpc (Mighell 1997). Moreover, the photometric
temperature as derived above was used in this relation, and the
stellar mass was taken to be 0.8$\,M_{\odot}$, in correspondence to
the typical main-sequence turn-off mass. The bolometric correction
(BC) was obtained by interpolation of the respective
Kurucz-atmosphere grids.

Adopting this set of values as the input parameters, the final
spectroscopic parameters were determined from a subset of the clean
Fe\,{\sc i} lines listed in Table~2, for which 
$-5.4\,\la\,\log_{10}({\rm EW}/\lambda)\,\la\,-4.5$, corresponding
to a range of $\sim$20 to 160\,m\AA\ at 5000\AA. The lower limit of
this selection prevents weak lines being spuriously detected,
thus artificially raising the mean iron abundance near the cut-off.
The strongest, saturated lines were pruned from the list,
since they do not lie on the linear part of the curve of growth and
are predominantly formed in the highest layers, where the atmosphere
models become progressively unreliable (McWilliam et al.\ 1995;
Johnson 2002). 

Atmospheric parameters were determined by iteratively modifying the
models to satisfy three constraints in successive steps. As a first
step, we determined the microturbulent velocity $\xi$.  For this
purpose abundances from the neutral iron lines were iteratively
calculated, demanding that there is no trend of abundance with the
reduced width EW/$\lambda$.  The scatter around the zero slope
from a least-squares fit was then taken as a measure of the
uncertainty on the microturbulent parameter. In this way, we
conservatively obtain typical accuracies of 0.1\,km\,s$^{-1}$ on
$\xi$.

In the second step, spectroscopic temperatures were obtained
by requiring excitation equilibrium.  Based on the range of
T$_{\mathrm{eff}}$\ that produces acceptable fits in the Fe\,{\sc
  i}-abundance versus excitation potential (EP) plot, we estimate the
typical errors of the spectroscopic temperature to be $\pm 100\,$K,
slightly larger than the photometry-based errors.  Generally, our
spectroscopic T$_{\mathrm{eff}}$-values are systematically higher than
the photometric estimates, and we note an offset of $+250$\,K on
average for our standard atmospheres.  For some of the targets this
discrepancy can be as high as 500\,K.  Although a systematic offset of
$\sim$250\,K may raise some concern, differences between the
spectroscopic temperature scales do not come as a surprise (see also
the discussion in Johnson 2002).  In our analyses we use the
spectroscopic temperatures to define the atmospheric
T$_{\mathrm{eff}}$\ for several reasons.  First, the photometric
errors of the infrared colors are rather high.  The median photometric
uncertainty of our targets from 2MASS amounts to 0.1\,mag in V$-$K. A
reddening uncertainty of 0.1\,mag leads to a T$_{\mathrm{eff}}$\
variation of $\pm$100K.  It turns out that an unreasonably large
systematic decrease of 0.4\,mag on average in V$-$K (up to 0.71\,mag
for the most deviant stars), be it due to photometric or reddening
uncertainties, would be required to reconcile the spectroscopic and
photometric temperatures.  However, this would correspond to an
unreasonably large increase of Carina's distance modulus of
0.45--0.8\,mag (reflecting a distance shift of $\sim$18--30\,kpc).
Inconsistent photometric and spectroscopic T$_{\mathrm{eff}}$-values
could occur due to errors in the $gf$-values, which may correlate with
the EP (Johnson 2002). However, this has been shown not to be the case
either for laboratory oscillator strengths (e.g., Blackwell et al.\
1995), nor should it be a problem in our differential line-by-line
analysis.  Finally, Johnson et al.\ (2006) have shown that the use of
a photometric T$_{\mathrm{eff}}$\ yielded highly diverging
non-standard abundance ratios for their sample of old red giants in
LMC clusters, again strengthening our confidence in the use of the
spectroscopic values.

In the third step, the appropriate surface gravity for each star was
constrained by assuming ionization equilibrium, thus requiring that
the log\,$g$-sensitive Fe\,{\sc ii} abundance  
 matches the abundance from the Fe\,{\sc i} lines, 
$\varepsilon($Fe\,{\sc i}). These abundances were repeatedly computed
for both species with log\,$g$ varying in small steps until the
results from both ionization stages agreed reasonably well. By
demanding that both ionisation levels do not deviate by more than 0.1
dex we could effectively constrain our gravity estimate to within an
accuracy of $\pm$0.2 dex.

As for the case of T$_{\mathrm{eff}}$, the gravities determined from
ionization equilibrium are also systematically higher than the 
photometrically derived values (eq. 1). This discrepancy cannot be resolved
by  adopting our spectroscopic T$_{\mathrm{eff}}$-values in eq.~1
instead of the color temperatures.  This would render the gravities
larger by $\sim 0.18$ dex on average, which is still lower than their 
spectroscopically derived values.
It is feasible that this is due to a mild degeneracy of the entire
parameter set, where a fraction of the offset can be explained by
discrepancies in the adopted temperature.  Since an increase in
T$_{\mathrm{eff}}$\ decreases the difference between
$\varepsilon($Fe\,{\sc i}) and $\varepsilon($Fe\,{\sc ii}), one
inevitably derives an increased stellar gravity.  Due to this
interdependence, a change in log\,$g$ entails a slight change in the
slope of the EP plot, which again is reflected by another rise in
effective temperature (e.g., McWilliam et al.\ 1995; Johnson 2002;
Johnson et al.\ 2006; Fulbright et al.\ 2006; see also Section 3.3).  Microturbulence also
has an influence on log\,$g$, though to a lesser extent.  
While matching $\varepsilon($Fe\,{\sc i}) and
$\varepsilon($Fe\,{\sc ii}) to achieve ionization equilibrium itself
can be done accurately, one should also keep in mind that there is only a
small number of stronger and well determined Fe\,{\sc ii} transitions
available in our spectra and in the literature.  On average, we have only
about four Fe\,{\sc ii} lines at hand, which will not only lead to larger
random errors on the derived [Fe\,{\sc ii}/H] values, but also
frustrate accurate gravity estimates. As we will show
in the next Section, even a implausibly large variation of log\,$g$
has an only marginal effect on the abundances, in particular since 
 we primarily rely on the neutral iron lines to
assign the stellar [Fe/H].

The entire parameter determination for $\xi$, T$_{\mathrm{eff}}$\ and
log\,$g$ was successively iterated to approach simultaneous
convergence, adopting the Fe\,{\sc i} abundance from the last
iteration as the input metallicity [$M$/H] (which strictly accounts
for the global metal abundance, e.g., Salaris et al. 1993) for the
subsequent atmosphere calculation.
The final atmospheric parameters for each star derived in this way are
tabulated in Table~3.  With this final optimized parameter set, the
abundance of the ionized state Fe\,{\sc ii} deviates by no more than
0.04 dex from the Fe\,{\sc i} abundance, with an average deviation of
$\sim$0.01--0.02 dex.

\subsection{Abundance errors}

In order to determine accurate
error estimates  in the individual abundance ratios, 
we investigated the effects of the random uncertainties in
the atmospheric parameters.  
First of all, abundance ratios were obtained for ten different model
atmospheres for two stars (LG04c\_000777 and LG04d\_006628).  These
two stars were chosen because of their significantly different stellar
parameters. 
In each
of the models, the stellar parameters were separately varied by their
typical random 1\,$\sigma$ uncertainty, as conservatively estimated in
the previous Section.  In Table~4 the differences of the 
abundance ratios for each of these models are given, i.e., for
T$_{\mathrm{eff}}$$\,\pm\,$100\,K, $\xi\,\pm\,$0.1\,km\,s$^{-1}$,
log\,$g\,\pm\,$0.2 dex and [$M$/H]$\,\pm\,$0.2 dex, w.r.t the values
obtained from the atmospheres computed with the best spectroscopic
parameter set.

At this stage of the analysis our model atmospheres for the Carina
stars incorporate opacity distribution functions assuming scaled-solar
element abundances ({\sc odfnew}).  It is important, however, to
keep in mind that the $\alpha$-elements are important electron donors
in the atmospheres of red giants.  These metals supply a significant
number of free electrons, which sustain the formation of H$^{-}$ --
the dominant source of the continuum opacity in these stars.  It
cannot be assumed {\em a priori} that the Carina stars exhibit any kind
of enhancement in these elements, but measurements in seven
dSphs of the Local Group, also comprising five stars in Carina (which
do not coincide with our sample), indicate that many stars in these
galaxies are in fact enhanced in a number of $\alpha$-elements
(Shetrone et al.\ 2001, S03; Sadakane et al.\ 2004; Venn et al.\ 2004;
Geisler et al.\ 2005).  Therefore we additionally derived abundances
under the assumption of an $\alpha$-enhancement of $+0.4$ dex in the
atmospheres, by calculating models with the stellar default
parameters, now using the $\alpha$-enhanced opacity distributions,
{\sc aodfnew} (see, e.g., Fulbright et al.\ 2006, 2007).  The change
in each elements' abundance ratio derived with this atmosphere from
the standard atmosphere's value is given in Table~4 in the column
labeled ``$\alpha$''.

The model parameters are not
independent but rather intricately correlated in several ways
(McWilliam et al.\ 1995).  This leads to covariances of the
parameters' uncertainties, which can be particularly significant for
the case of T$_{\mathrm{eff}}$\ and log\,$g$ caused by an 
explicit temperature-gravity dependence.  A change in
T$_{\mathrm{eff}}$\ can also introduce a slope in the abundance vs. EW
relation, which is used to fix $\xi$. 

To quantify the covariances of the stellar parameters, we follow the
prescription of Johnson (2002), which is, in turn, based on the
formalism of McWilliam et al. (1995). For the representative stars
LG04c\_000777 and LG04d\_006628 we start by drawing 20 random
temperature values from a Gaussian with their spectroscopic
T$_{\mathrm{eff}}$\ as mean and with a dispersion of 100\,K. For each
of these, log\,$g$ was changed accordingly, to reinstall ionization
equilibrium. The resulting covariance
\begin{displaymath}
\sigma_{\rm T\,log\,g}=1/N\sum_{i=1}^{N}{(T_i-<T>)(log\,g_i-<log\,g>)}
\end{displaymath}
amounts to 40, with a correlation coefficient of 0.79, i.e., the
mutual influences of temperature and gravity on our abundances are in
fact highly correlated.  Likewise, we computed 20 values for the
microturbulence with a dispersion of 0.1\,km\,s$^{-1}$ and readjusted
the gravity of the atmosphere to regain ionization equilibrium. These
parameters are less correlated, with $\sigma_{\rm \xi\,log\,g}=-0.005$
and a correlation coefficient of 0.44.  It turns out that covariance
terms including the atmospheres' metallicity [$M$/H] are negligible so
that we did not include the respective mixed terms in the analysis.

Finally, we account for random errors in the EW measurements by
computing the 1$\sigma$ scatter in the abundances, weighted by the
square root of the number of absorption features, $N$, used in the
derivation of $\varepsilon$ (e.g., Carretta 2006). This statistical
error, $\sigma_{\rm EW}/\sqrt{N}$, (see also Table~5) accounts in
principle for the line-to-line scatter and input atomic parameters.
For those elements, for which only one absorption feature was
measurable in the spectra, we adopted a representative error on the
abundance through repeatedly varying the EW in the abundance
determination within its typical accuracy (of the order of 5--10\%).
In general, for those elements with $N\ga5$, which is the case for our
Ca-, Ti\,{\sc i}\,- and Fe-lines, the total abundance error is
dominated by the uncertainties in the atmospheric parameters
(Prochaska et al.\ 2000).  The major error source in terms of
atmospheric parameters is in this case the temperature sensitivity.
For elements with few lines, there is a progressively larger
contribution of EW errors due to enhanced noise, unresolved blends or
uncertainties in the continuum placement.  Interestingly, neglecting
$\alpha$-enhancement does not significantly alter any of the derived
abundance ratios and only has a significant effect on the ionized
species Ti\,{\sc ii} and Fe\,{\sc ii}, where the enhancement tends to
overestimate the abundances by $\sim$0.1 dex.

The total uncertainty $\sigma_{\rm tot}$ of each elemental abundance
ratio [X/H] due to random errors in the atmospheric parameters (listed
in Table~5) is then obtained by summing the uncorrelated contributions
from each individual parameter (T$_{\mathrm{eff}}$, $\xi$, log\,g,
[$M$/H], $\alpha$, $\Delta$EW) in quadrature, and adding the
respective covariant contributions from T$_{\mathrm{eff}}$, $\xi$ and
log\,g (see eq.~2 in Johnson 2002).  Since variations in the
atmospheres will affect different elements to a different extent, one
has to additionally consider their covariances in computing the
abundance ratios [X/Fe]. Therefore we obtain $\sigma^2_{\rm
  [X/Fe]}=\sigma^2_{\rm X}+\sigma^2_{\rm Fe}-2\sigma_{\rm X\,Fe}$ with
the covariance $\sigma_{\rm X\,Fe}$ according to eq.~6 from Johnson
(2002).
As a result, the typical error on our derived [\ion{Fe}{1}/H] amounts
to $\sim$0.17 dex and ranges from $\sim$0.05 to $\sim$0.20 dex for the
other elements, while the uncertainty on [O/Fe] can be as high 0.35
dex.

Finally, we note that none of our abundances has been corrected for
NLTE effects.  Since our results will be compared below with similar
LTE analyses (e.g., S03; Venn et al.\ 2004), this source of
uncertainty will not be systematic on the whole.  Despite our
enforcement of ionization equilibrium by reconciling
$\varepsilon$(\ion{Fe}{1}) and $\varepsilon$(\ion{Fe}{2}), the neutral
and ionized species of Ti show a marginal difference of $-0.12\pm0.18$
dex. 
If we demanded ionization equilibrium for the Ti
lines to establish a log\,$g$, our spectroscopic surface gravities
would have had to be lower by 0.13 dex on average, which is well
within our estimate of the uncertainty on this atmospheric parameter
(see Section 3.2).  The fact that our data do not exhibit
any significant trend of $\varepsilon($Ti\,{\sc i}) or
$\varepsilon($Ti\,{\sc ii}) with T$_{\mathrm{eff}}$\, nor with metallicity,  supports our view
that we can, for the moment, neglect NLTE effects in our analyses
(Luck \& Bond 1985; cf. Johnson 2002).

\section{Results}

Table~5 shows the final abundance ratios [$X$/Fe] ([Fe/H] for the iron
lines themselves) for each chemical element $X$. 
These ratios are given w.r.t.\ the mean abundance of
the neutral iron lines, $\varepsilon$(Fe\,{\sc i}).  Also tabulated
are the number of lines, N, used to determine the unweighted mean
abundance ratio for each element and standard deviation of each
$\varepsilon$ measured  from these lines.

\subsection{Iron}

The mean iron abundance of the red giants in Carina obtained from our
high-resolution spectra is  $-$1.69 dex with a standard
deviation of 0.51 dex.  This is in excellent agreement with the mean
of $-$1.72 (on the scale of Carretta \& Gratton 1997) derived from
the CaT measurements.  Also S03 find a mean of $-1.64$ with a scatter
of 0.2 dex from their high-resolution analysis of five red giants,
while Smecker-Hane et al.\ (1999) reported a slightly lower value of
$-1.99$ dex from a small number of stars with CaT measurements.
{From} this point of view, our targets can be considered as Carina
members representing the typical metallicity range in this dSph.

There are some mildly metal-rich targets, which reach [Fe/H] of
$-1.27$ dex.  Furthermore, we detect two metal-poor stars with
[Fe/H]$<-2.5$ dex, which contribute to the metal-poor tail of Carina's
MDF.  These stars lie 1.5--2\,$\sigma$ below the peak of the
MDF. Nevertheless, this small number of metal-poor objects is not
sufficient to resolve the apparent G-dwarf problem (or in fact
``K-giant problem'', Koch et al.\ 2007) in dSphs (Smecker-Hane et al.\
1999; Pont et al.\ 2004; Tolstoy et al.\ 2004;
Venn et al.\ 2004; Helmi et al. 2006;  Paper\,I;  
Bosler et al.\ 2007; Koch et al. 2007). Interestingly, at 1.4 and 2 core radii, these two stars are
also located at the largest radial distances in our sample, thus
matching the mild radial abundance gradient that was detected in
Paper\,I (see also Harbeck et al. 2001). Recall that our parent sample
selected for spectroscopy was three times larger than the eventual
member sample, and was selected to ensure that any very metal-poor or
metal-rich member star would be included. The absence of such stars
from our sample then indicates a real scarcity.
\subsubsection{Assessing the CaT calibration}
The six stars that were observed in both low- and high resolution mode
enable us to assess the accuracy of the calibration of the CaT
abundance scale onto iron, since these stars cover a range of
abundances. Fig.~1 shows a comparison of the estimates from both
methods.

As the dashed line of unity in Fig.~1 (top panel) implies, all of the
five more metal-rich red giants show good consistency between their
CaT metallicity (on the scale of Carretta \& Gratton 1997) and the
high-resolution iron abundance, not deviating by more than
0.8\,$\sigma$ from unity.  This range of metallicities is well sampled
by the Galactic globular clusters (GGCs) in the calibrations of
Rutledge et al.\ (1997a,b).  An oft-cited caveat against
calibrating the CaT in dSphs to a {\em Galactic} reference system is
that GGCs tend to exhibit enhanced [Ca/Fe] ratios of $\sim +0.4$ dex
on average (e.g., Pritzl et al. 2005), while {\em dSph} stars show
lower values in this ratio (Venn et al. 2004).  As we will show in
Sect. 4.3, our Carina dSph stars exhibit slightly lower [Ca/Fe] ratios
of the order of $+0.09 \pm 0.14$ dex, which should result in
systematically lower line strengths of the CaT compared to the GGCs.
A number of clusters in the Rutledge et al. (1997b)
sample show in fact lower [Ca/Fe] values of $\sim +0.1$ dex at
metallicities similar to that of our Carina stars (e.g., Cohen \&
Mel\'endez 2005), and the GGCs, from which we derived the calibration of the CaT
to [Fe/H]$_{\rm CaT}$ in Paper\,I, possess an average, enhanced,
[Ca/Fe] of $+0.21\pm 0.11$ dex.
Thus we might expect an inferred [Fe/H]$_{\rm CaT}$ lower by $\sim
0.12$ dex with respect to the [Fe/H] abundance, based on the
difference in [Ca/Fe] between target and calibrators (see also
discussions in Paper\,I;  Bosler et al. 2007).  We find that our CaT
based measurements are more metal poor by (0.08$\pm$0.09) dex on
average. This small zero point shift is consistent with the above
difference in [Ca/Fe] of the calibration clusters and target stars.
Note that there is no obvious trend of the difference between
high-resolution [Fe/H] and [Fe/H]$_{\rm CaT}$ with [Ca/Fe] discernible
in our data.

The iron abundance of our metal-poor star is, at $-$2.50 dex, lower by
0.44 dex (2.1\,$\sigma$) than the CaT value.  This deviation is in
qualitative agreement with the CaT vs. high-resolution [Fe/H]
comparison of a large number of dSph stars of Battaglia et al. (2008),
who show that a systematic overestimate of the CaT abundances sets in
below [Fe/H]$\la -2.2$ dex.  Such a trend points to a general
shortcoming of the method to reliably infer iron metallicities 
from the CaT index in the most metal poor red giants.
Since the globular clusters in the sample of Rutledge et al.\ (1997b)
``only'' comprised a metallicity range from $-$0.64 down to $-$2.02
dex (Carretta \& Gratton [1997] scale), measurements of stars beyond
these limits necessarily rely on an extrapolation of the calibrations.
In particular, the clusters employed for our calibrations in Paper\,I span
a range from $-2.0$ to $-1.12$ dex.

As an additional test, we calibrated our GGC measurements from
Paper\,I onto the \ion{Fe}{2} scale of Kraft \& Ivans (2003). The
major difference to the old calibration lies in the [Fe/H] value for
the cluster M\,68, for which Kraft \& Ivans (2003) find a more metal
poor [Fe/H] of $-2.43$ dex, while this cluster has a metallicity of
$-2$ dex on the previously used scale of Carretta \& Gratton (1997).
In general, Carretta \& Gratton (1997) have reported the most metal
poor GGCs to be considerably more metal rich than other studies.  The
main impact of the new calibration (see bottom panel of Fig.~1) is
that the most metal poor of our targets becomes even more deficient,
at [Fe/H]$_{\rm CaT} \sim -3.1$ dex, which underestimates the
metallicity compared to our high-resolution measurement by 0.56 dex.
On average, the CaT measurements on the Kraft \& Ivans (2003) scale
are more metal poor by 0.3 dex than the UVES data, with a standard
deviation of 0.12 dex around unity.
Although Cole et al.\ (2004) and, more recently, Carrera et al. (2007), 
have demonstrated that the CaT technique
is generally applicable for a broad age range of 0.25--13\,Gyr and
across a wide metallicity range spanning $-2.2<$[Fe/H]$<+0.47$, the MDFs'
metal-poor tails remain unsatisfactorily sampled in most high
resolution studies.  The question of whether the metal-poor extensions
of the MDFs of dwarf galaxies based on the CaT below $\sim -2.2$ to
$-2.5$ dex, which are reported throughout the literature (e.g.,
Smecker-Hane et al.\ 1999; Tolstoy et al.\ 2001; Bosler et al.\ 2007;
Pont et al.\ 2004; Tolstoy et al.\ 2004; Cole et al.\ 2004, 2005;
Paper\,I) are then reliable has to await more and highly accurate data
in this regime (e.g., Battaglia et al.  2008).  Nonetheless, the CaT
index is well suited to the identification of candidates for fairly
metal-poor stars, although the exact [Fe/H] value may not be
measurable -- this is then left for follow-up high-resolution studies.
\subsection{Alpha Elements -- O, Mg, Si, Ca, Ti}
While iron is synthesized both in SNe\,II and SNe\,Ia, the even-$Z$
elements (O, Mg, Si, Ca, Ti) are predicted to be produced during 
(nucleosynthetic) shell-burning in high-mass stars with very short
lifetimes, which later end their lives in SN\,II explosions. 

A simple scenario for the formation of the Galactic halo is that an
initial burst of SF produces large amounts of
$\alpha$-elements on short time scales, leading to an enhanced
[$\alpha$/Fe] ratio.  At these early epochs, the long-lived SNe\,Ia
did not yet have time to significantly pollute the initial SNe\,II
abundance pattern.  Once SNe\,Ia start to contribute (after a delay of
$\sim$1\,Gyr from the onset of SF), iron production will
increase, leading to a decline in the [$\alpha$/Fe] abundance
ratio. As noted above, the detailed patterns of the transitions in
[$\alpha$/Fe] with metallicity (taken as a proxy for time) are
dependent on the specific SF history of a stellar system,
on its stellar IMF, and on the timescales for mixing the SNe products
into the interstellar medium (ISM) (Matteucci 2003; S03; Venn et al.\
2004).

The metal-poor stars of the Galactic halo evidently populate a plateau
of [$\alpha$/Fe]$\sim$0.4 dex (see, e.g., the review by McWilliam 1997).
From the available analyses of 36 red giants in seven dwarf
galaxies, S03 found at best little evidence to identify these stellar
populations with those in the Milky Way.  In fact, stars in dSphs have
been shown to be quite different from Galactic disk, and even more so
from halo stars, in that they have lower [$\alpha$/Fe] ratios (e.g.,
Shetrone et al.\ 2001; Pritzl et al.\ 2005; see also Fig.~3).  Hence,
these elements have often been considered as an important 
test for Searle \& Zinn's (1978) hypothesis of the origin of the
Galactic halo from satellite accretion events (Unavane, Wyse and
Gilmore 1996).

Despite the characteristic relatively low [$\alpha$/Fe] in most stars
of the dSphs, their wide abundance range from approximately $-3.0$ to
0.3 dex and the fact that some stars also exhibit higher, Galactic
[$\alpha$/Fe] values suggests that using these elements as a test for
chemical imprints of accretion events requires a sophisticated
analysis (Venn et al.\ 2004). Irrespective of this analysis, the
$\alpha$ elements provide important information in elucidating the
evolutionary histories of individual dwarf galaxies (e.g., S03; T03).
We will turn to discussing the implications of our derived
[$\alpha$/Fe] ratios for Carina's chemical evolution in Section~5.

Figs.~2 through 4 display the derived abundances for each
$\alpha$-element in our Carina stars, together with the measurements
of five red giants in Carina of S03, where none of their targets
coincides with our present sample.  The $\alpha$-elements are
displayed individually as well as a combination of these elements    as 
the parameter [$\alpha$/Fe]=[(Mg+Ca+Ti)/3\,Fe] (see also Table~1 in
Venn et al.\ 2004).  The right panels of Fig.~2 and Figs.~3, 4 also
include a compilation of Galactic thin and thick disk stars as well
as a large number of halo field stars\footnote{The stellar abundances
  for these illustrations are tabulated in Venn et al.  (2004), who in
  turn assembled the data from various literature sources. See their
  paper for references.}.  Note that the separation into Galactic
components is purely based on kinematic information.

In order to compare Carina's elemental abundances with those of stars
in other Local Group galaxies, we show also currently available
abundance ratios for stars in eight Galactic dSphs (Draco, Sextans,
UrsaMinor from Shetrone et al.\ 2001; Sculptor, Fornax, Leo\,I, Carina
from S03; UrsaMinor from Sadakane et al.\ 2004; Sculptor from Geisler
et al.\ 2005 and Sagittarius from Monaco et al. 2005).

When comparing data sets from various sources, as was done by Venn et
al.\ (2004) and as is shown in Figs.~3
and 4 here, one has to keep in mind that each analysis may use
different spectral lines, presumably differing oscillator strengths,
and different model atmosphere techniques. The combination of such
data will then result in an analysis-dependent spread in the abundance
ratios, which is of the order of 0.1--0.2 dex (Venn et al.\ 2004), a
value comparable to the abundance offsets found by S03 by using
different stellar atmospheres and atomic data.  In the light of our
measurement uncertainties (Sect.~3.3; indicated by the error bars in
the figures), we did not attempt to homogenize the data shown in the
subsequent Figures. This measurement method induced scatter does not
alter any of the conclusions we draw below.

The first thing to note is that one of our target stars,
LG04a\_001826, exhibits a strong depletion in almost all of the
$\alpha$-elements, although it is moderately metal rich, at
[Fe/H]=$-$1.49 dex.  This star will be discussed further below.

As Venn et al.\ (2004) note, the scatter in [Mg/Fe] is expected to be
much larger than that in either [Ca/Fe] or [Ti/Fe], an expectation
which is met by our data.  Apart from the strongly discrepant target
at [Fe/H]=$-$1.49, with [Mg/Fe]=$-0.92$, the more metal rich targets
cover a full range in [Mg/Fe] from $-$0.24 to $+$0.08 dex, while the
two most metal-poor stars in our sample exhibit enhanced Mg abundances
of 0.3 and 0.4 dex, and thus partially overlap with the Galactic halo.
In order to test the extent to which the scatter can be considered
intrinsic, we compared the observed scatter in the abundance ratios to
the value predicted from a set of Monte Carlo simulations, in which
our measurements were varied by the observational uncertainties. While
the predicted and observed $1\sigma$-scatter are similar for the cases
of [Ca/Fe] and [Ti/Fe], the observed scatter in [Mg/Fe], at 0.16 dex,
exceeds that expected from the pure observational uncertainties by
$\sim$0.05 dex, indicating a real dispersion.

Ca is predominately produced in intermediate-mass progenitor
SNe\,II (Woosley \& Weaver 1995).  
While the most metal-poor stars of the Shetrone et al.\ (2001) and S03
samples in Sextans and Draco are clearly depleted w.r.t.\ the Galactic
components (at [Ca/Fe]$\sim$0.1), our most metal-poor Carina giant is
significantly enhanced in Ca (by 0.7 dex). However, already the second
most metal-poor star (0.2 dex more metal rich) has a [Ca/Fe] that is
only slightly larger than the average we find from the rest of our
sample ($<$[Ca/Fe]$>$\,=\,0.09$\pm0.14$ dex).  While the majority of
our targets with [Fe/H]$>-$2 tend to follow the trend of the other
dSphs' Ca abundances, and also are consistent with S03's measurements,
one should note that three or four of the more metal-rich giants have
comparable enhancements to those of the Galactic thick disk and/or
halo.

Although it is still unclear if the nucleosynthesis of Ti proceeds in
analogy to the other $\alpha$-elements (e.g., Timmes et al.\ 1995;
Woosley \& Weaver 1995), it is traditionally included in the group of
these elements due to its similarly enhanced abundances in metal-poor
stars (Gratton \& Sneden 1991). 
With the exception of the strongly depleted LG04a\_001826, Ti
in Carina tends to follow the overall trends of the other
$\alpha$-elements, with an average abundance significantly lower than
the Galactic halo star abundance ratios.  In fact, the two most
metal-poor stars of our sample show enhanced [Ti/Fe] of $\sim$0.1 and
0.36 dex, which is consistent with the Galactic halo, but at odds with
the two comparably metal-poor red giants from Shetrone et al.\ (2001)
and S03. The latter stars, in the Sextans and Draco dSphs, are
deficient in [Ti/Fe] by $\sim-$0.23 dex.  The majority of our targets
is consistent with both the Carina data and most of the other dSphs'
abundances of Shetrone et al.\ (2001) and S03.

Si is often considered  a prototypical $\alpha$-element. Its
production site is identified with SNe\,II of moderate progenitor
masses, around 20\,M$_{\odot}$ (Woosley \& Weaver 1995). In our data
we were able to determine its abundance from typically three
lines. For the most metal-poor star no information could be extracted,
since no measurable line was detected in its spectrum.

In the second most metal-poor star, there is a strong indication of a
significant enrichment in Si, as found in most metal-poor stellar
abundance studies (McWilliam 1997; Prochaska et al.\ 2000; Venn et
al.\ 2004, and references therein). As Fig.~4 implies, an enhancement
of $\sim$0.9 dex is not atypical for metal-poor giants in dSphs and
also is found in UMi (Shetrone et al.\ 2001). Our stars show a lower
average [Si/Fe] compared to the Shetrone samples, where our mean value
is $<$[Si/Fe]$>$\,=\,$-0.01\pm0.12$ dex (for [Fe/H]\,$>-$2). It is
worth noticing, however, that there is a fair amount of scatter in
[Si/Fe], both in the Galactic halo and all dSph samples considered
here, including our own data.  In this context, [Si/Fe] abundance ratios 
as low as $-0.4$ are also found in other dSphs of the Local Group.
While the second most metal poor star
is enhanced in [Si/Fe], the abundance ratios around Carina's mean
[Fe/H] typically reach from about $-$0.2 up to +0.15 dex in our
targets.  Owing to the overall scatter, some of the dSphs' stellar
abundances overlap with the Galactic halo component at [Si/Fe]$\sim
+0.4$.  

Oxygen abundances were measured from the two forbidden lines at
$\lambda\lambda$6300, 6363 \AA, which are, however, not detectable in
the most metal poor target, nor in the overall depleted star of
LG04a\_001826.  We find a range in [O/Fe] from $-$0.2 to $+$0.65 dex
with a considerable star-to-star scatter at a given [Fe/H] (Fig.~5,
top panels).  As Fig.~5 (bottom) implies, the range in
[O/Fe] in Carina, as well as the increased scatter compared to the
other $\alpha$-elements, is consistent with that found in other
dSphs. As mentioned above, these abundances are depleted w.r.t. the
Galactic halo by $\sim$0.5 dex and show a well defined decline with
[Fe/H] at lower metallicities than in the Galaxy.

\subsection{Sodium} Sodium abundances were determined from the
absorption lines at 5682 and 5688\AA.  We did not include hyperfine
splitting in the calculations, since its effect is expected to be
negligible (McWilliam et al. 1995).  As Fig.~5 (bottom panels)
implies, our mean [Na/Fe] in Carina of $-0.36$ dex with its scatter of
$\sim$0.27 dex is consistent with the Carina values of S03, and also
with the [Na/Fe] in the other dSphs.  In particular, there is a strong
overlap of Carina's Na abundance with the Galactic halo stars, which
also cover a broad range.

It is worth noticing that there is no trend of [Na/Fe] with oxygen
 discernible. Such a pattern would be evidence of deep mixing.  In deep
 mixing, processed material from the interior of a star can be
 transported into the upper atmosphere.  Nucleosynthetic burning
 converts O, N and Ne to Na, whereas Mg is converted into Al in the
 deeper atmospheric layers of red giants. Hence, typical deep mixing
 patterns predict that an atmospheric depletion in O coincides with an
 enhancement of Na (Langer et al. 1993), while any observed Mg
 depletion is accompanied by an enhanced Al abundance (Langer \&
 Hoffmann 1995).  Unfortunately, in none of our spectra could Al be
 detected, since all suitable lines in our spectral range
 are too weak.  Interestingly, such (anti-)
 correlations are detected in globular clusters both in the Galaxy
 (Pritzl et al. 2005 and references therein) and in dSphs (Letarte et
 al.\ 2006), but they are not observed in Galactic field stars
 (Gratton et al.\ 2004) nor in dSph field stars (S03; Venn et al.\
 2004).  Consistent with this, our Carina red giants show a constant
 mean [Na/O] of $\sim -0.63\pm0.17$ over the full range of
 metallicities.

\subsection{The $\alpha$-depleted case of LG04a\_001826} %
We now turn
to the question of whether our target star LG04a\_001826, which shows
a considerable depletion in almost all of the $\alpha$-element ratios,
i.e., in [Mg/Fe], [Ca/Fe] and [Ti/Fe], can provide further insight
into general galactic evolution or whether it should rather be
considered a special case. For this star [Mg/Fe] lies 2.2$\sigma$
below the sample average, and [Ca/Fe] and [Ti/Fe] are underabundant by
1.8$\sigma$ and 2.3$\sigma$, respectively, w.r.t. the mean abundance
ratios in Carina.

One possible interpretation of a selective depletion of the
$\alpha$-elements could be the occurrence of deep mixing, but as
discussed above, this is not seen in other stars in dSphs. Moreover,
at [Na/Fe]=$-0.48$ LG04a\_001826 shows a sodium abundance that is
consistent with the mean [Na/Fe] ratio found from all our Carina
stars, and  Na is not overly enhanced w.r.t. the non-detection of oxygen in
its photosphere.  It appears unlikely that this star's abundance
patterns are influenced by deep mixing.  An untestable hypothesis is that 
enrichment by rapidly rotating massive stars may be involved, since this can 
lead to globular-cluster-like abundance anomalies, dredging-up processed
material into the outer layers of the massive stars, where it may
be lost as a wind prior to the Type II supernova  (Decressin et al.\
2007).

Generally speaking, the $\alpha$-elements are synthesized in
core-collapse SNe\,II, where 
different elements originate in different stages of these endphases in
the lives of massive stars.  Oxygen and magnesium, which show the
strongest depletion (Mg) or upper limits (O) in this star, are
produced mainly during the hydrostatic burning phases.  Their yields
are primarily a function of the progenitor's mass, but are heavily
affected by the thickness of the burning shells, the treatment of
rotation and convection, and the mass cut between the proto-neutron
star and the SN ejecta (Thielemann et al.\ 1996; Argast et
al. 2002). If one assumes a larger range of possible realizations of
these factors, a larger scatter in the resulting element ratios such
as [O/Fe] and [Mg/Fe] is feasible (Argast et al.\ 2002; Venn et al.\
2004). Depending on the exact mass of the SN\,II progenitor, a low
value of [Mg/Fe] is possible (see Fig.~1 in Argast et al.\ 2002). Low
[Mg/Fe] is observed in a few metal poor bulge-stars (McWilliam et al.\
1995) and in mildly metal-poor halo stars (e.g., Carney et al.\ 1997;
Ivans et al. 2003).

Some SN models (like those of Argast et al.\ 2000, 2002) predict the
occurrence of such depleted stars at lower, halo-like metallicities,
specifically due to the mixing of SN ejecta into a gas reservoir of
$\sim10^4$M$_{\odot}$. Comparable variations at a higher [Fe/H], as
seen in the case of LG04a\_001826, can be explained
analogously with reasonable parameters, particularly noting the
uncertainties associated with the modelling of Mg and Fe yields.
Alternatively, one could assume that our target star exhibits a normal
average abundance of Mg, but has an anomalous Fe abundance.
Then its low [Mg/H] would be representative of a halo-like star of
[Fe/H]$\sim-2.6$ dex. In this case, in order to realize its
comparatively low [Mg/Fe], LG04a\_001826 had to originate from SNe\,Ia
pre-enriched material (Gilmore \& Wyse 1998).  This would be diluted
less than the characteristic halo ISM or could be subject to a
significantly higher iron yield, producing a [Fe/H] higher by
$\sim1.1$ dex than expected from its $\alpha$-yield.

The other $\alpha$-elements Si, Ca and Ti originate in explosive
burning during the SN event itself and depend primarily on the
explosion energy and to a lesser extent on the amount of material
available to be burned and the binding energy of the star, that is,  
its mass (e.g., Woosley \& Weaver 1995).  Since typical
explosion energies cover only a small range, the abundance ratios of
these elements can be expected to show a much lesser scatter, which is
in fact in agreement with the observations (Venn et al. 2004; see also
Figs.~3 and 4).

It is then odd that this star exhibits a rather average [Si/Fe] ratio (at
0.2$\sigma$ below the sample mean).  It has been suggested that Si
contains a non-negligible admixture of material originating in
hydrostatic burning (Timmes et al.\ 1995; Thielemann et al.\ 1996;
Argast et al.\ 2000) and it is then likely that the resulting
abundance patterns can differ from a pure explosive origin as for Ca
or Ti (but see Cohen et al. 2007 for the exact opposite pattern).

The intermediate-mass $\alpha$-element Ti has
also a contribution from complete Si burning with an $\alpha$-rich
freeze-out (Thielemann et al.\ 1996 and references therein).
Differences between Ca, Si, and Ti are thus naturally to be expected.
Finally, the production of Ti is not yet fully understood (Woosley
1986; Woosley \& Weaver 1995) and Galactic evolution models often fail
to reproduce the observed abundance patterns (e.g., Timmes et al.\
1995).
A detailed interpretation of this star's abundance ratios in terms of
the yields for each individual $\alpha$-element has to rely
on fine-tuned nucleosynthetic modeling  and is presently hampered by
the lack of an accurate understanding of the exact SN\,II explosion
mechanism (e.g., see discussions in Woosley \& Weaver 1995; Timmes et
al.\ 1995; McWilliam et al.\ 1995; Thielemann et al.\ 1996; Argast et
al.\ 2000, 2002). 
  
All in all, the presence of a strongly $\alpha$-element depleted
object would seem to suggest that we see here the influence of {\em
single} SN events. Contrary to a homogeneous, close-to-uniform
[$\alpha$/Fe] pattern resulting from well-mixed, IMF-averaged SNe,
this star is likely to reflect the imprint of stochastic inhomogeneous
chemical evolution, where the star formed from unmixed SN ejecta.
Rather than (over-) interpreting this star's abundance pattern in terms
of Carina's SF history we attribute it to statistical fluctuations due
to incomplete ISM mixing.  That such incomplete mixing may occur also
in actively star-forming dwarf galaxies was recently shown by Kniazev
et al.\ (2005).  It is then also possible to assign the strong Ca
enhancement of the most metal-poor star in our sample to such scatter,
which becomes progressively dominant in metal-poor stars of
[Fe/H]$\la-3$ dex. This contrasts with the low scatter in the more
metal poor Galactic halo, indicative of a well-mixed ISM even at very
early times (e.g., Cayrel et al. 2004; Fran\c cois et al. 2004).  Of
course, any distinction between Carina and the halo based on the small
number of peculiar stars in Carina should be taken with caution: while
the dispersion in the abundance ratios of metal poor halo stars is in
fact smaller than would be expected if {\em all} metal poor stars were
polluted by single SNe, there are a number of chemically peculiar
metal poor stars present in the halo (e.g., Norris et al. 2002; Cohen
et al. 2007).

Fortunately, observational analyses of $\alpha$-depleted stars in the
Galactic halo are underway, and will quantify how SNe\,Ia and SNe\,II
contribute to their abundance patterns.  For instance, Ivans et
al. (2003; and references therein) investigate three moderately metal
poor halo stars ([Fe/H]$\sim -2$) with unusually low $\alpha$- and
neutron capture element abundances.  By comparison with the abundance
patterns of very metal poor stars, these stars were argued to be iron
rich rather than $\alpha$-poor.  As in our Carina case, their
depletion in [$\alpha$/Fe] would be explicable in terms of a
contribution of SNe\,Ia yields that is larger than expected given
their low metallicities.  It is conceivable that such stars may have
been born out of material that was polluted by the earliest SNe\,Ia to
have occurred in the Milky Way (Nissen \& Schuster 1997; Ivans et
al. 2003).  It has been suggested that at least a fraction of this new
class of Galactic halo stars may have been captured from accreted,
dSph-like systems.  Despite the general discrepancy between the halo
and present-day dSph abundances, our finding of such a peculiar,
$\alpha$-depleted star in Carina then makes feasible a partial
accretion origin of this chemically distinct halo subpopulation.

\section{Implications for Carina's chemical evolution}

Apart from the two most metal-poor stars in our sample and the
strongly depleted red giant LG04a\_001826, the majority of our targets
are compatible with solar $\alpha$-abundances: we find a mean
[$\alpha$/Fe] of 0.00$\pm0.10$ dex. Our data confirm the earlier
important finding that the $\alpha$-elements in dSph stars are
significantly depleted w.r.t.\ Galactic halo stars of the same
metallicity (Shetrone et al.\ 2001; S03; T03; Venn et al.\ 2004).  
It is interesting to note that the two most metal-poor stars in Carina
 show an enhancement in the $\alpha$-elements by $\sim 0.3$ and 0.4
 dex, which is the same order of magnitude as seen in the halo stars.
 These stars should then also be expected to be the oldest stars to
 have formed and may have been born at an {\em early} epoch, when
 SNe\,Ia had not yet started to contribute, but when an initial
 starburst had already polluted the pristine material with a significant amount of the $\alpha$-elements ejected in short-lived SNeII.  In general,
 dSph stars populating the same SNe\,II plateau as the Galactic halo
 are consistent with an invariant, massive star, IMF (e.g., Wyse et
 al. 2002).

On the other hand, those stars with higher metallicities (between
$\sim-1.8$ and $-1.3$ dex) resemble stars in other dSph
galaxies.
 These progressively lower values of
[$\alpha$/Fe] are characteristic of the slow chemical evolution and
low SF rates in dwarf galaxies (Unavane et al. 1996; Matteucci 2003).
Differences between individual galaxies' abundance ratios are,
of course, to be expected, since these chemical tracers sensitively
depend on the details of each galaxy's environment, SF history, IMF,
etc.  Overall, our stars' abundance ratios are also in good
agreement with the values found in the five stars studied by S03.

It has been proposed that Carina's episodic SF history, i.e., its
sequence of bursty and quiescent phases, is reflected in the
[$\alpha$/Fe] abundance ratios (S03, T03). S03 suggested that the
considerable rise in [$\alpha$/Fe] -- after the typical decline due to
the continuing feedback from SNe\,Ia -- seen in their data at
[Fe/H]$\sim-1.65$ indicates that these stars may be related to either
Carina's second SF episode around 7\,Gyr ago or to an
intermediate SF event well before this episode. Both interpretations
are compatible with photometric determinations of Carina's SF history
and the times of each SF episode and hiatus (Smecker-Hane et al.\
1996; Hurley-Keller et al.\ 1998; Rizzi et al.\ 2003). Any connection
is strengthened by comparable variations in a number of heavy element
abundance ratios. Resolving and quantifying the SF history clearly
requires further investigation: the interpretation of S03 and T03 was
based on only 5 stars with limited age resolution. 

We combine our data for the $\alpha$-elements O, Mg, Si and Ca with
the measurements from S03. The resulting data set represents the
entire currently available high-resolution abundance data for Carina
(we do have a larger study underway).  It is tempting to
interpret the [$\alpha$/Fe] pattern of the full data set (Figs.~3 and
6) in terms of Carina's episodic SF.  S03 have noted a rise in
[$\alpha$/Fe] between $-1.65$ and $-1.6$ dex from their 5 stars.  This
rise is also discernible in the full data set including S03's and our
ten stars, where it is most distinct in [O/Fe] and less pronounced in
the other elements.  After a decline in SF rates, the [$\alpha$/Fe]
ratio may rise again towards a maximum at around
[Fe/H]=$-1.4$ dex, before leveling off again.  However, in the light
of the measurement uncertainties (see also Fig.~6, bottom right) the
run of [$\alpha$/Fe] vs. [Fe/H] for most of our stars (disregarding
the most metal-poor and strongly $\alpha$-depleted star for the
moment) and those of S03 can equally well be considered to be constant
(see bottom right panel of Fig.~6), with some, perhaps intrinsic,
scatter in [$\alpha$/Fe]. Uniformly low [$\alpha$/Fe] ratios over a
wide range of metallicities would then point to the fact that Carina's
SF was in fact dominated by SNe\,II driven blow-out (e.g., Mac Low \&
Ferrara 1999; Paper\,I).

In Fig.~7, we compare the full observational set with the models from
 Lanfranchi et al.\ (2006, hereafter LMC06), which are an extension of
 the models presented in Lanfranchi \& Matteucci (2004).  This set of
 models was tailored, modifying best-fit SF and wind efficiencies, so
 as to reproduce the total mass, present-day gas mass and the few
 elemental abundances available at that time for a number of dSphs.
 In addition to the aforementioned observables used in Lanfranchi \&
 Matteucci (2004), LMC06 also incorporated our MDF of Carina from
 Paper\,I to update their choice of model parameters. A marginally
 higher SF efficiency (of 0.15\,Gyr$^{-1}$) and a slightly lower wind
 efficiency (of 5 times the SF rate) were adopted to better match our
 observations.  The models described in Lanfranchi \& Matteucci
 (2004) assumed Carina's SF history according to Hernandez, Gilmore,
 \& Valls-Gabaud (2000), and so are lacking a significant very old
 population, (e.g., Smecker-Hane et al. 1994; Harbeck et al.\ 2001;
 Monelli et al. 2003; Grebel \& Gallagher 2004). The LMC06 models
 adopt the SF history from Rizzi et al.\ (2003). In this latter study,
 the SF was found to have occurred episodically with an initial active
 period at an early epoch, while the majority of stars were formed
 $\sim$3--7\,Gyr ago.

 For all the elements shown in Fig.~7 there is a sudden decrease in
the computed [$\alpha$/Fe] at [Fe/H]$\sim-1.6$, consistent with a
combination of the onset of galactic winds and the injection of Fe
into the ISM via SN\,Ia events.  Once the winds set in, the available
amount of gas for SF is reduced, resulting in a decline of the overall
SF. Consequently, the formation of new stars almost ceases and, in
turn, the occurrence of SNe\,II and the related production and
injection of the $\alpha$-elements is halted as well.  Since 
Fe is continuously produced and injected into the ISM by
the longer timescale SNe\,Ia for several Gyr, unaffected by cessation
of SF, the start of efficient winds leads to the sudden decrease seen
in the modeled [$\alpha$/Fe] abundance ratios.  It is, however, not 
 obvious {\em a priori} that the SF rate and thus the associated SNe\,II rate
was high enough to sustain winds -- these free parameters are put in
by hand in the LMC06 models.

For Ca the majority of our stars lie above the models, while the
observed [Mg/Fe] and [Si/Fe] ratios tend to lie below the predictions
at any given metallicity.  This may in part be attributed to the large
scatter in [$\alpha$/Fe] for basically all metallicities, a scatter
which also has been reported to exist in previous observational
studies (see Pritzl et al.\ 2005 and references therein).  At present,
the models appear to provide at best an upper limit to the
observations, given the overall large star to star scatter,  
and thus reproduce the more chemically enhanced stars
reasonably well. 
One should keep in mind
that the original model parameters were optimized to fit the abundance
patterns as observed by S03 so that the overall trends in their
element ratios are fit fairly well.

The overall impression to glean from Fig.~7 is that, in
spite of being able to predict certain patterns in chemical abundance
ratios, there is a general need for models of chemical evolution to
steadily improve the parameters governing the SF processes.  In
essence, matching the full sample of available $\alpha$-elemental data
appears to require a turnoff from the plateau in [$\alpha$/Fe] at {\em
lower} metallicities (again with the exception of Ca).  These low
abundances at higher metallicities may be achieved by adopting a less
intense wind and/or a cessation in SF at earlier epochs. More intense
winds  would provide lower values for {\em all}
[$\alpha$/Fe] ratios, once the wind developed, due to the
correspondingly decreased SF rate.  Limitations can be attributed to
the difficulties in assuming a specific SF history for the models and
will depend both on the number, duration, and times of the assigned SF
episodes.  Similarly, an overall lower SF efficiency would be able to
retain the lower [$\alpha$/Fe] values.  Since these parameters affect
the shape of Carina's MDF (recall that they have been adjusted to match the
latter) it is clear that ideally the MDF, the full set of observed
abundance ratios and a fully sampled SF history have to be
simultaneously taken into account in order to reliably devise the
evolutionary history of Carina.  Considering the discussion in
Sect. 4.4, it is also conceivable that stochastic SF, e.g., in small
associations, and inhomogeneous enrichment, may have played a major
role in Carina's history.

In Fig.~8 we overplot a set of chemical evolution models from Gilmore
\& Wyse (1991) with our observed oxygen abundance ratios. In these
models, which were the first chemical evolution models to account 
naturally for below-solar $\alpha$-element ratios at low
metallicities, 
SF occurs in an initial, short-lived burst, during which there is no
significant contribution from SNe\,Ia.  This active period is followed
by a quiescent phase lasting several Gyr, during which only SNe\,Ia
enrich the ISM. A second, late SF burst is assumed to set
in at higher metallicities. In the figure, the first burst occurs at
[Fe/H]=$-$2.1 dex, while the second SF sets in at $-1.1$ (solid line)
and $-1.4$ dex (dashed line). We did not aim at optimizing the exact
onset of each burst to constrain Carina's SF history, but rather
shifted the predictions of Gilmore \& Wyse (1991), which were
originally computed to represent the chemical evolution of the LMC and SMC. It is
then intriguing that already this simple scenario yields an amazingly
good representation of our data, and that the observed increase and scatter in the
abundance patterns can in fact be assigned to an increase in SF
activity, perhaps associated with separate SF episodes 7 and 3 Gyr ago
(Hurley-Keller et al.\ 1998).  This comparison indicates that the
onset of the second epoch of SF has to be around an
abundance of $-$1.5 dex.  It cannot be ruled out by our data that two
SF bursts have occurred in the interval between $\sim -2$ and $\sim
-1.4$ dex, which is probable given the fact that deep photometric
studies agree on the occurrence of three major SF episodes in Carina
(Smecker-Hane et al. 1996).

The occurrence of SF bursts should cause variations in both [O/Fe] and
the other $\alpha$-elements, whereas the observed scatter in [Ca/Fe] and
[Ti/Fe] is smaller than that seen in our [O/Fe] abundance ratios.  As
Gilmore \& Wyse (1991) argue, inhomogeneities in the location and
onset of each SF burst inevitably leads to a large scatter in the
abundance ratios, such as seen in Carina.  Even in the presence of a
statistically universal IMF, the intrinsically rare high-mass stars
will form with significant random number fluctuations in SF events of
small total mass:  SF events that produce only a small total mass in stars might not form any of the
highest-mass oxygen-producing stars at all.  Adopting a standard IMF
(e.g., Miller \& Scalo 1979), it can be shown that $M_{\rm tot}\sim
100M_{\odot}$ form in low-mass stars per SN\,II ($M\ge10 M_{\odot}$),
while $\sim 4000 M_{\odot}$ form in the lower-mass regime of the IMF
for every 40$M_{\odot}$ star.  Thus, in a system like Carina, where SF
involves small numbers, as well as low rates, the resulting
mass-averaged IMF yield will have inevitable stochastic scatter,
contributing to observed intrinsic scatter in our abundance ratios.  
 Effective truncation of the massive-star IMF due to a low
SF rate was suggested by T03 as a mechanism to produce low
values of [$\alpha$/Fe] in dSph stars, treating the entire galaxy as a
``single-cell star-forming entity''. Our model comparison and the new, more numerous data allow us
here to identify additional new key elements -- inhomogeneous, stochastic
SF in low-mass associations or clusters, with incomplete
mixing leading to inhomogeneous chemical evolution. 
Overall, our data suggest that Carina's evolution involved
non-monotonic, stochastic, low-rate SF that forms 
marginally bound low-mass stellar associations and/or clusters, convolved with
incomplete mixing of the SNe ejecta into the ISM.

\section{Conclusions}

We have derived elemental abundance ratios for a number of light
chemical elements in ten red giants in the  Carina dSph,
tripling the existing sample for this galaxy.  By adding these ten
stars to the presently existing data for $\sim$50 red giants in eight
out of the 17 currently known dSph satellites of the MW system, the
sparse chemical information available for dSph galaxies slowly
continues to grow, confirming the overall trends in element abundance
ratios noticed in earlier studies.

As a number of our stars were also observed at lower resolution, we
could compare metallicity estimates from the well established {\em
calcium} triplet (CaT) calibration with the high-quality {\em iron}
abundances from the present work. We show that the CaT-based
metallicities and the true iron abundances agree remarkably well in
the relatively metal-rich regime of \mbox{[Fe/H]$\ga-$2 dex}.  For the
single more metal-poor star overlapping with the  high-resolution sample we find a
substantial difference in the metallicity derived from the CaT and
high-resolution iron lines. This may be since the CaT calibration
needs to be extrapolated below [Fe/H]$\la -2$ dex.  Since we have only
one data point, a more complete sampling of the dSphs' metal-poor
stars with high-resolution spectroscopy is desirable.  The red giants
in Carina span a comparable range in [Ca/Fe] to the GGCs 
that we used to calibrate Carina's low-resolution MDF in
Paper\,I, so we argue that the application of the CaT method to devise
a global metallicity is valid within the range of the calibration.

The majority of the elements observed follow the patterns seen in
other MW satellite dSphs (S03; Venn et al.\ 2004), i.e., they lie
generally below the Galactic halo star values.  This is consistent
with low SF efficiency governing the internal evolution of
the dwarfs, but is no consistent with halo formation by accretion of
present-day satellites (Shetrone et al. 2001). Our Mg abundances
show larger scatter than do Ca and Si, as  anticipated by their
different nucleosynthetic formation processes.

We detect two metal-poor objects at $-$2.5 and $-$2.7 dex. These are
significantly $\alpha$-enhanced and exhibit halo-like abundance
ratios, while the most metal-poor star is even more strongly enhanced
in Ca. Overall, it appears that the Ca ratios are slightly higher on
average than the remainder of the $\alpha$-elements.  Therefore the
relatively metal poor stars in the Galactic halo can in fact be
considered as remnants from an accreted dSph satellite, if this merger
event occurred very early on (e.g., Grebel et al. 2003; Carollo et al. 2003).

Our sample hosts one star that is remarkably depleted in almost all
elements.  Possible explanations include a particularly high iron
yield or the imprint of single SN events, the latter of which points
to the presence of early stochastical fluctuations and an incomplete
mixing of the ISM gas.  This may indicate that SF in Carina
occurred stochastically within smaller constituents, followed by an
inhomogeneous enrichment of the surrounding medium.  If SF in
dissolving star clusters was a major contributor in Carina, one might
be able to identify the clusters' remnants. Chemical substructure is,
however, generally not observed in the low mass dwarf galaxies (but
see Kleyna et al. 2003), nor in transition-type galaxies.  Localized
evidence of abundance imhomogeneities on small scales at a given time,
which argues in favor of SF in {\em associations}, has been detected
in a few dwarf irregular galaxies (e.g., Grebel \& Richtler 1992;
Gonzalez \& Wallerstein 1999; Kniazev et al. 2005).

We compare our data to models that include several bursts in Carina's
SF history and conclude that the anticipated low SF efficiencies 
and moderate mass-loss rates in dSphs provide reasonable
representations of present-day data sets, although they fail to
produce the observed abundance scatter.  There are still deficiencies
in the detailed models concerning some single elements. In particular
the Ca measurements are at odds with the models, having systematically
higher abundances on average.  We note that the LMC06 models that we
compared to our data did not account for incomplete mixing, which we,
however, observe.  By adopting a model prediction with two SF bursts
and an extended hiatus, which is qualitatively justified by Carina's SF history 
from its CMD, we argue that the second burst in Carina must
have occurred at an approximate [Fe/H] of $-$1.5. This burst
metallicity is not quantitatively constrained by CMDs, and hence is a prediction for
future abundance measurements (see also S03).  Such predictions become
testable, once metallicities and robust age estimates of stars on the
subgiant branches of Carina are available.
Although our data do not rule out any imprints of Carina's episodic
SF, we did not attempt to quantify the underlying exact SF history, which
is further impeded by the lack of accurate age information for our target
stars. 

\acknowledgments

A.K.\ and E.K.G.\ are grateful for support by the Swiss National
Science Foundation (grants 200020-105260; 200020-113697).  M.I.W.\ acknowledges the
Particle Physics and Astronomy Research Council for financial support.  

We would like to thank A. McWilliam for providing his versions of the
atmosphere interpolators and photometric gravity codes and helpful
discussions.  F.-K. Thielemann and G. Lanfranchi are acknowledged for helpful comments.

\newpage
\begin{figure}
\begin{center}
\includegraphics[width=0.7\hsize]{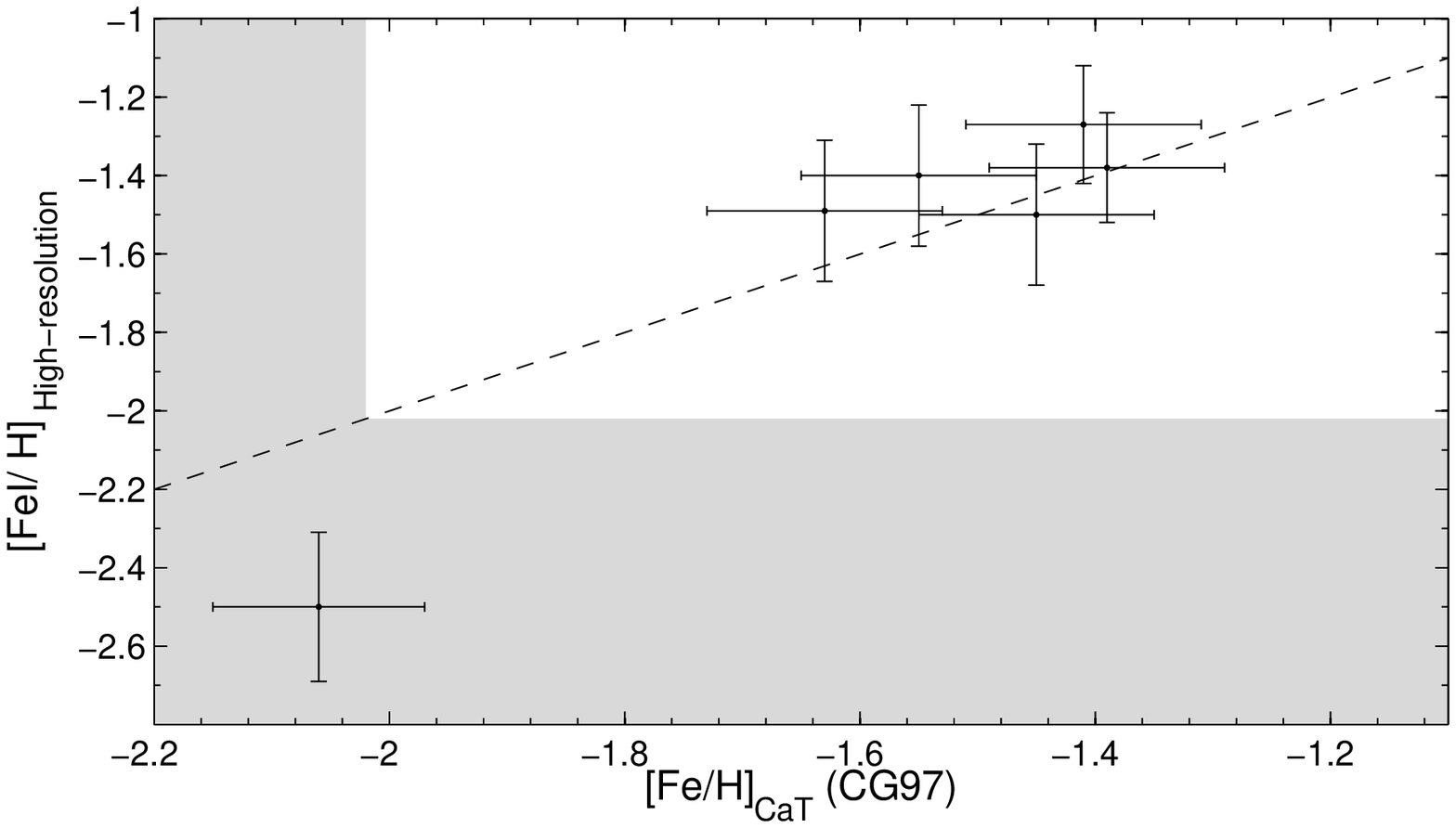}
\includegraphics[width=0.7\hsize]{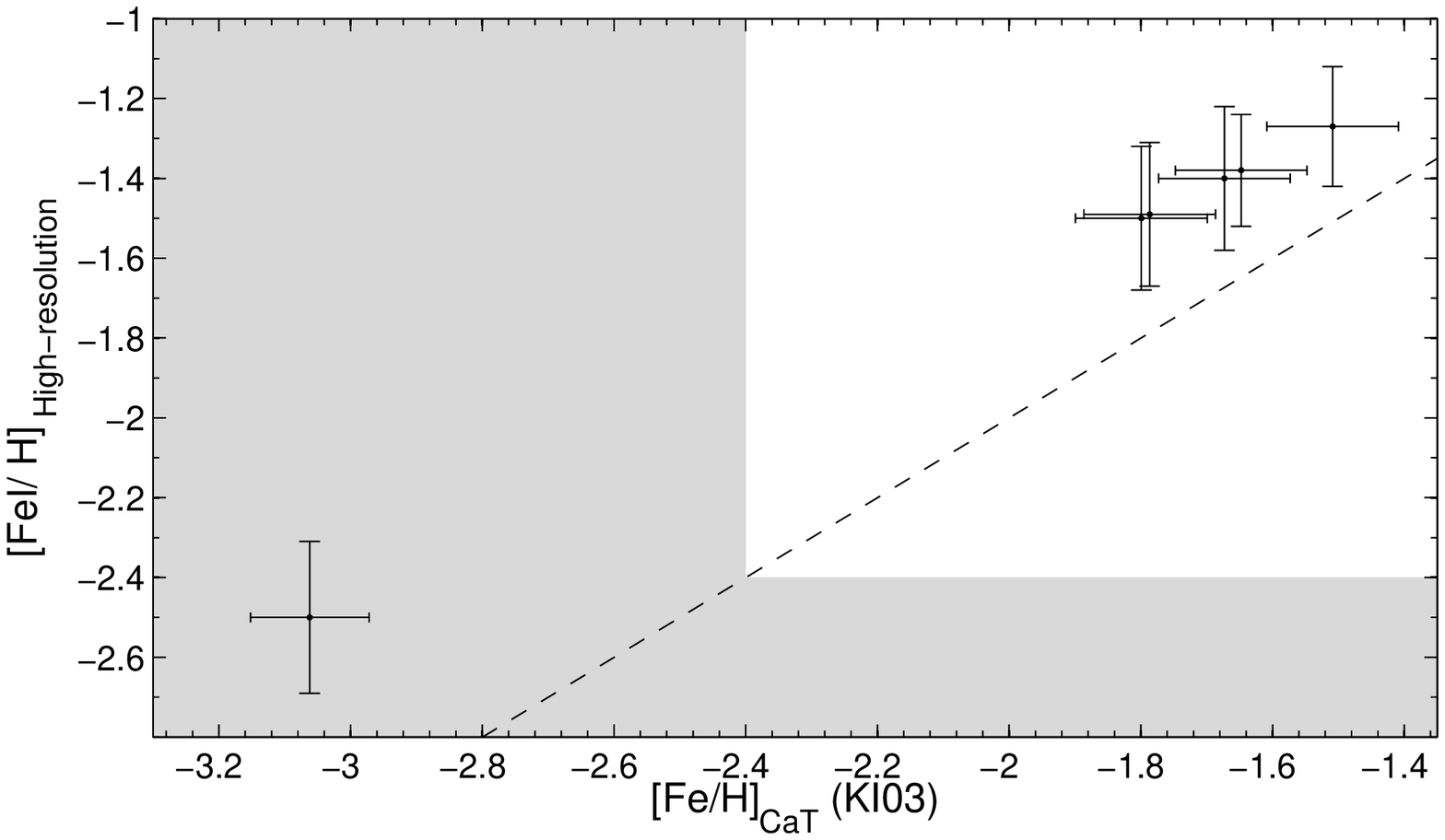}
\end{center}
\caption{Comparison of the metallicities from high- and medium-resolution (CaT) analyses 
on the metallicity scales of Carretta \& Gratton  (1997; CG97; top panel) and 
Kraft \& Ivans (2003; KI03; bottom panel). 
Shown are those six of our target stars, for which CaT widths were  measured in Paper\,I. 
The horizontal errorbars are formal measurements errors. 
The 1\,$\sigma$-errorbars on 
[Fe\,{\sc i}/H] are those deduced in Section 3.3. The grey-shaded areas illustrate the ranges, where 
the CaT calibrations have to be extrapolated due to the lack of very metal poor calibrating clusters in Rutledge et al.  (1997a,b) and Kraft \& Ivans (2003).}
\end{figure}
\begin{figure}
\begin{center}
\includegraphics[width=0.45\hsize]{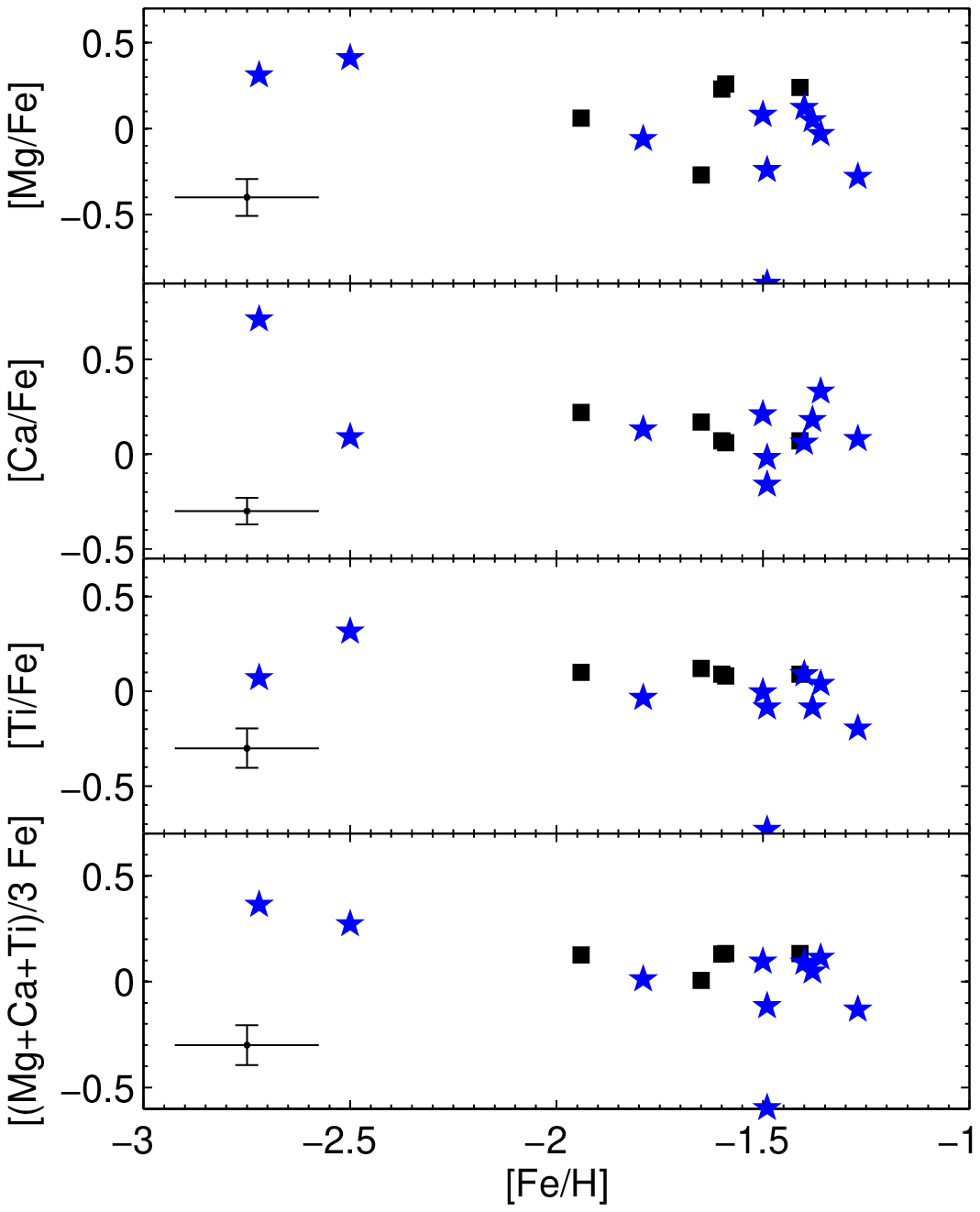}
\includegraphics[width=0.45\hsize]{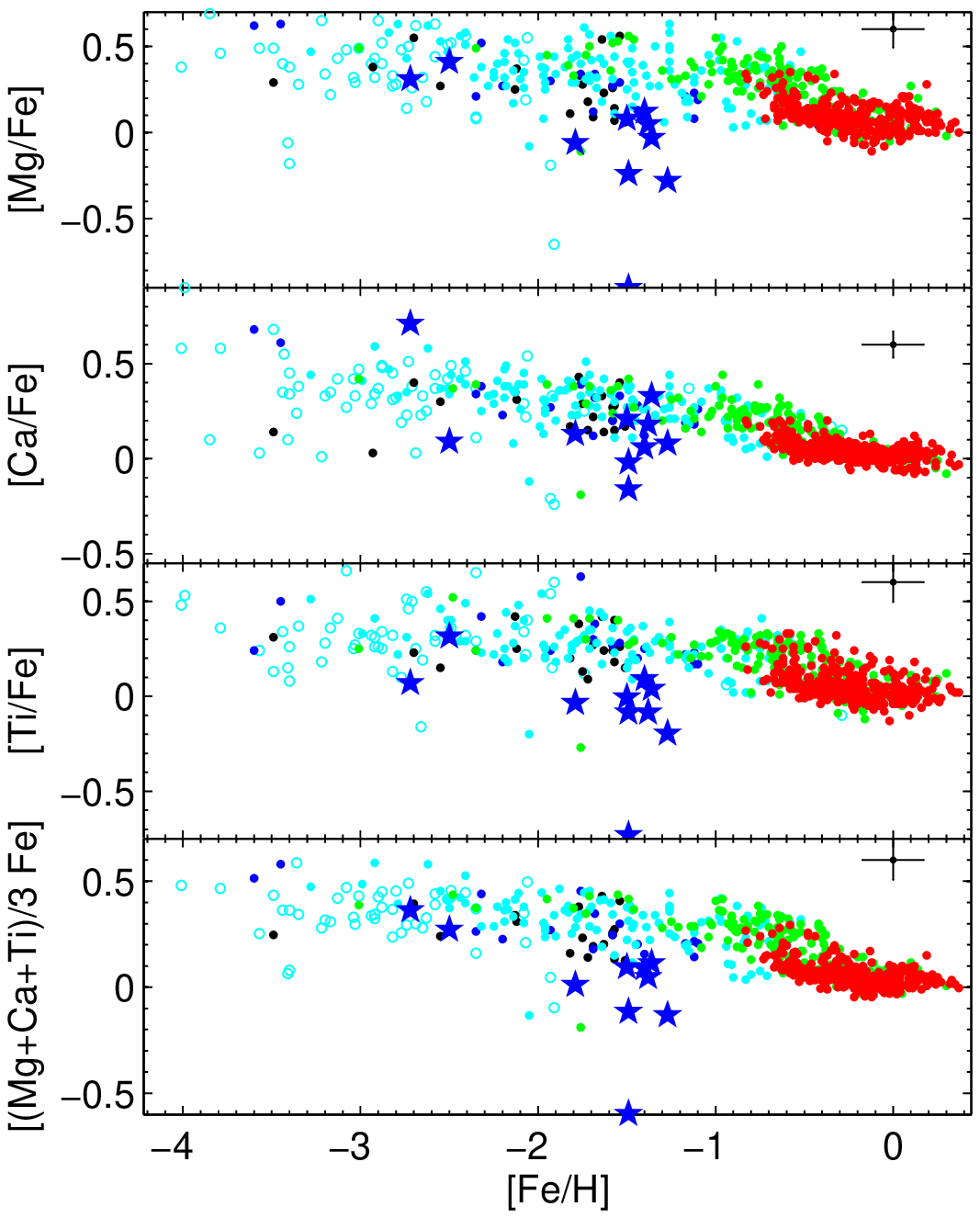}
\end{center}
\caption{Abundance ratios for the $\alpha$-elements Mg, Ca and Ti: 
Our measurements for the ten Carina red giants are shown as filled blue stars,  the 
filled black squares in the left panels are the Carina stars from S03. 
The additional data points in the right panels are Galactic field stars, where  
the color coding illustrates the kinematic separation into Galactic components and is shown 
in analogy to Venn et al. (2004); halo (cyan), thick  disk (green), thin disk (red), retrograde orbits 
(black) and those with a high-velocity Toomre component (blue). For stars shown as open cyan 
symbols, there is no existing velocity 
information. The errorbars in the upper right corner  are those derived in Section~3.3. 
[{\em See the electronic edition of the Journal for a color version of this figure.}]}
\end{figure}
\begin{figure}
\begin{center}
\includegraphics[width=1\hsize]{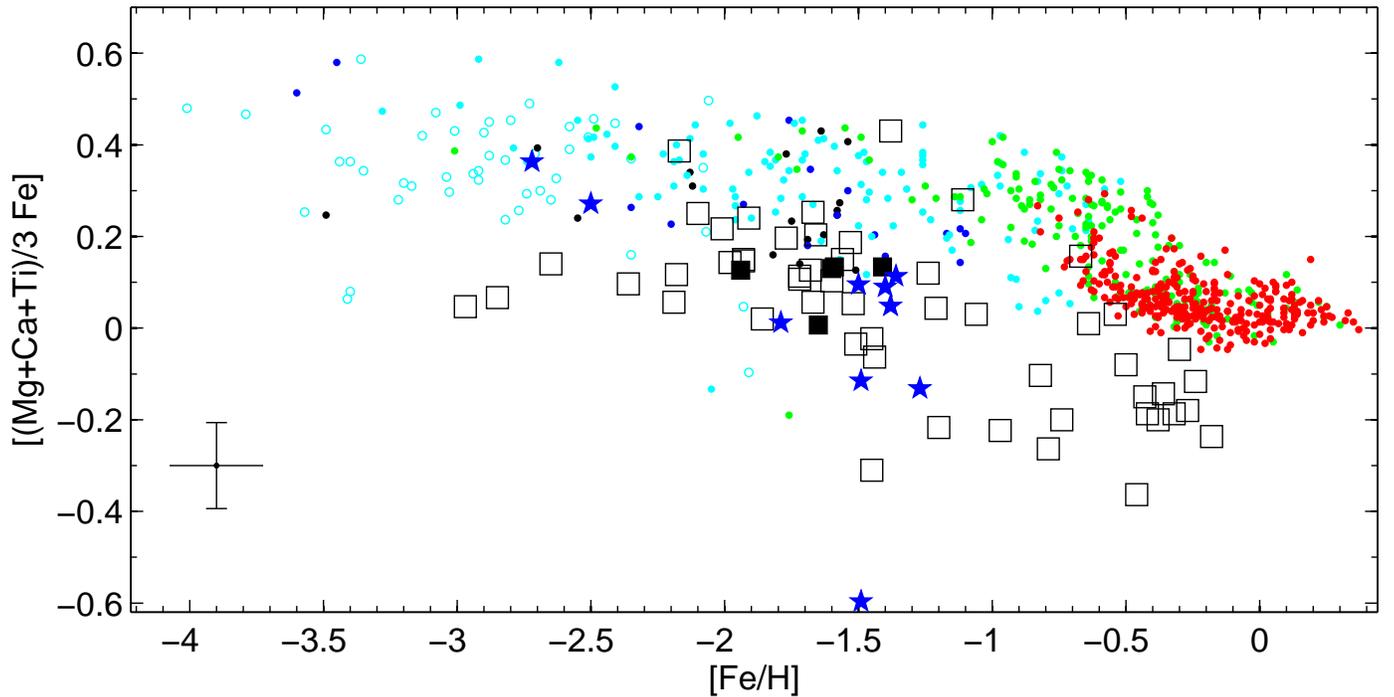}
\end{center}
\caption{Average $\alpha$ abundance ratio in analogy to Fig.~2. In addition, chemical element data 
of stars in eight other dSphs are shown as open squares (Shetrone et al. 2001, S03; Sadakane et al. 2004; Geisler et al. 2005; 
Monaco et al. 2005). 
[{\em See the electronic edition of the Journal for a color version of this figure.}]}
\end{figure}
\begin{figure}
\includegraphics[width=1\hsize]{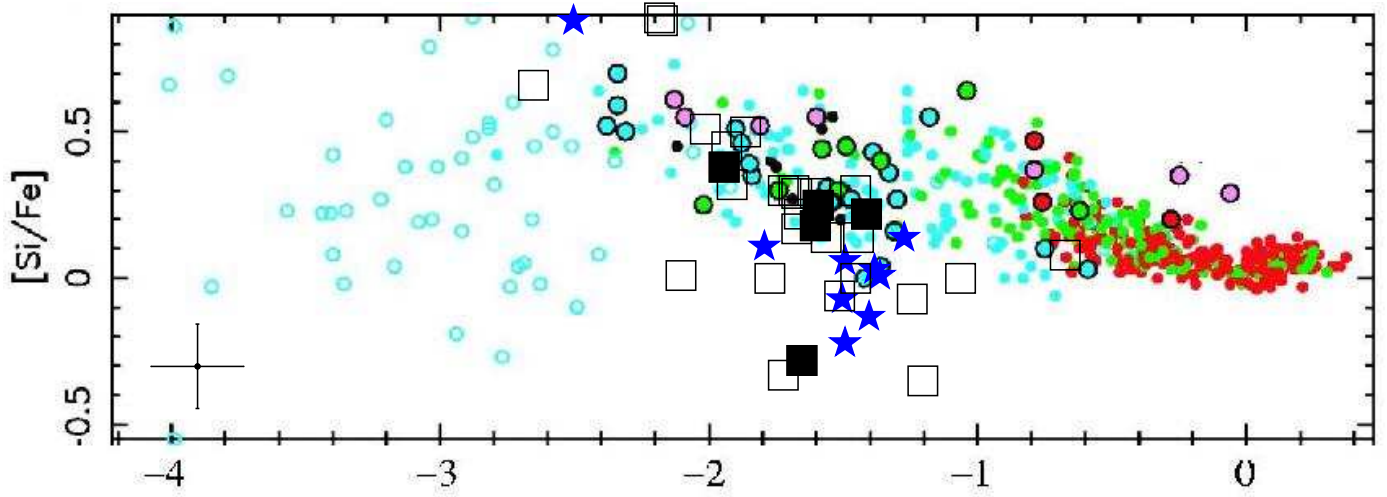}
\caption{Same as Fig.~3, but for silicon  abundances. Due to the lack of Si information in Venn et al. (2004), 
this figure was taken from Pritzl et al. (2005) and
additionally displays a sample of globular clusters (filled large circles). Reproduced with kind permission by 
B.~J. Pritzl. 
[{\em See the electronic edition of the Journal for a color version of this figure.}]}
\end{figure}
\begin{figure}
\begin{center}
\includegraphics[width=0.75\hsize]{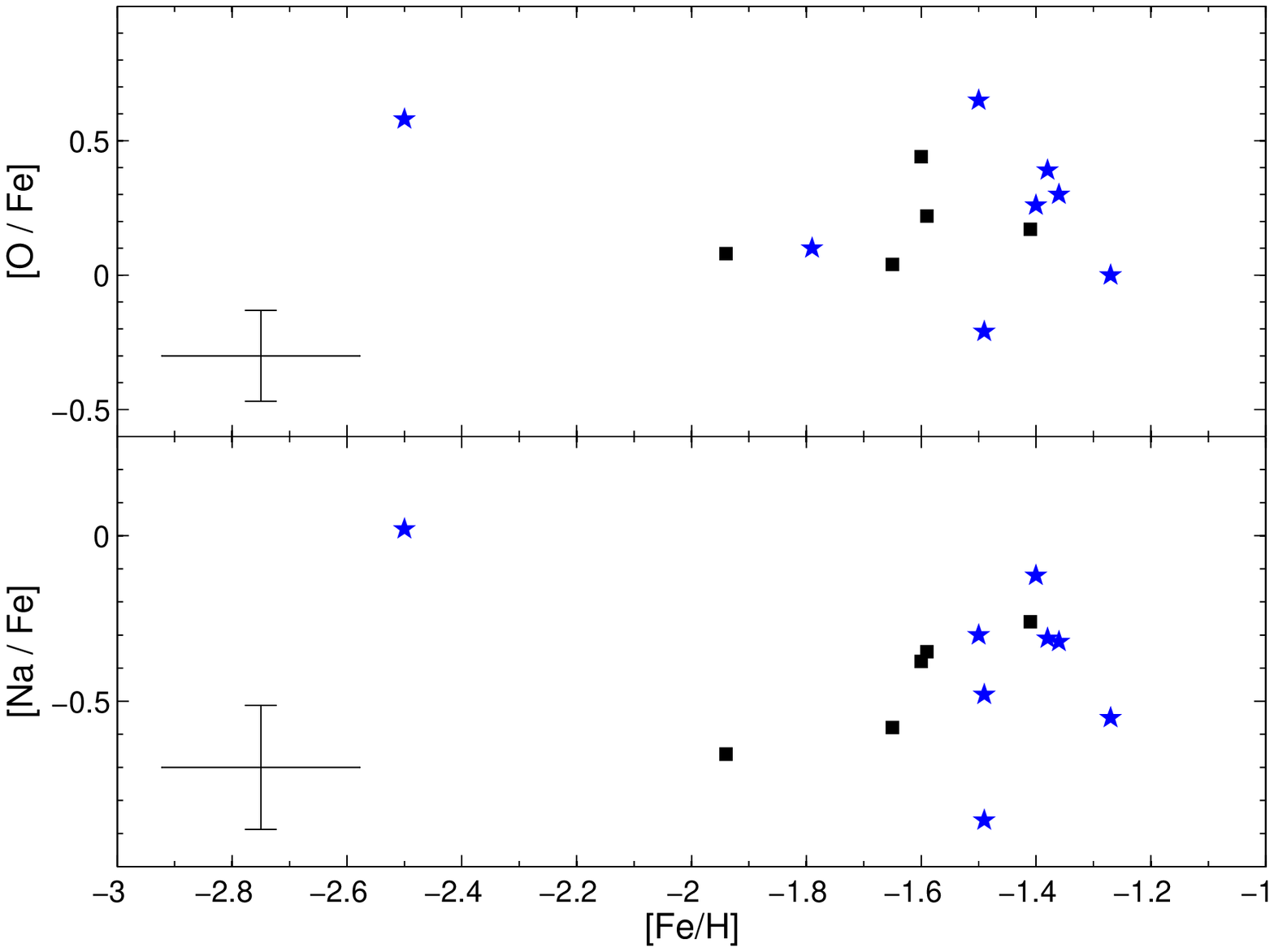}\\
\includegraphics[width=0.75\hsize]{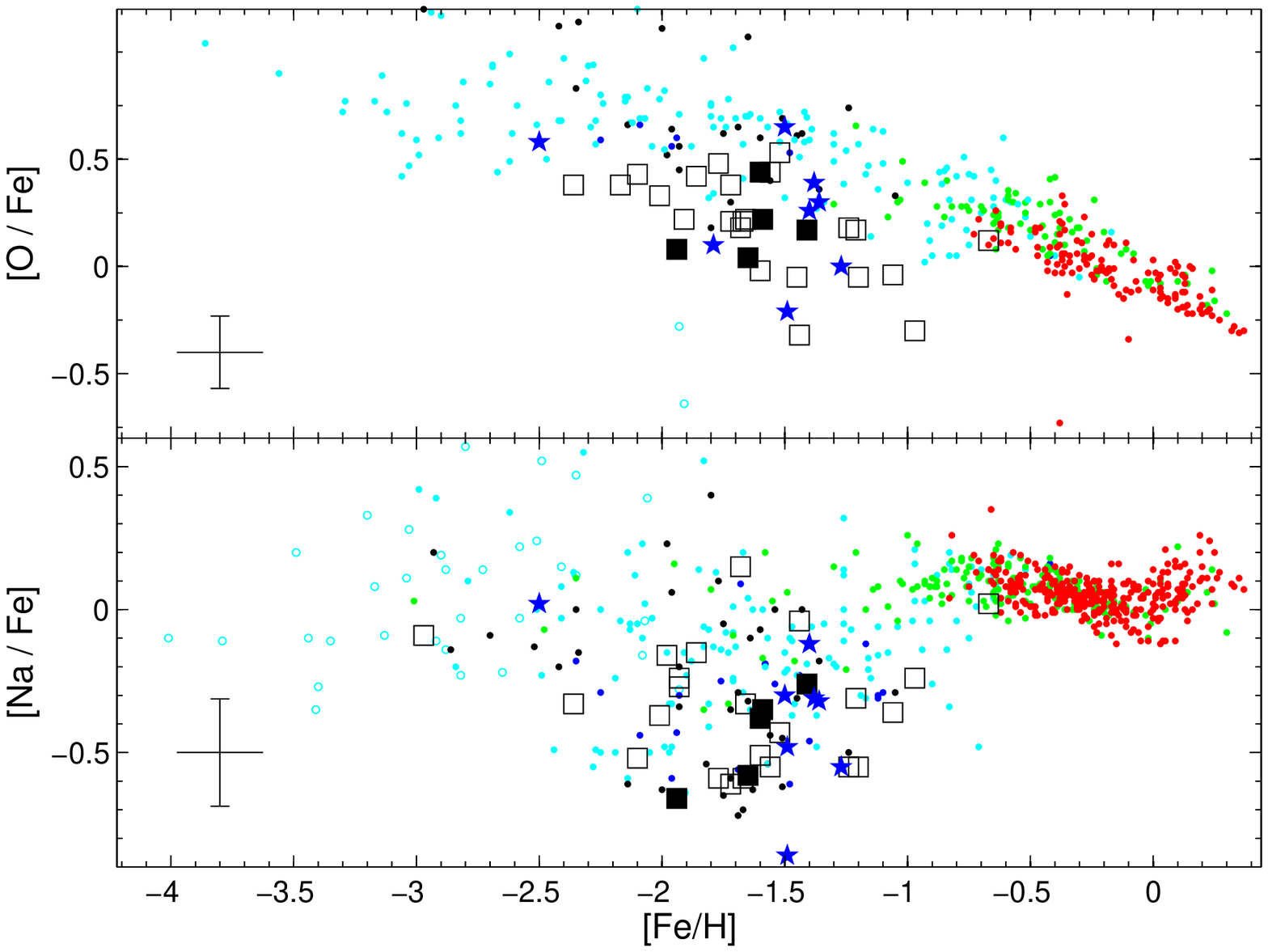}
\end{center}
\caption{Same as Fig.~2, but for Na and O  abundances. For illustration, the oxygen data 
were complemented with halo stars from the samples of Israelian et al. (1998);  Boesgaard et al. (1999);  
Mishenina et al. (2000); Akerman et al. (2004); Bai et al. (2004); Cayrel et al. (2004) and Zhang \& Zhao (2005). 
[{\em See the electronic edition of the Journal for a color version of this figure.}]}
\end{figure}
\begin{figure}
\begin{center}
\includegraphics[width=14cm]{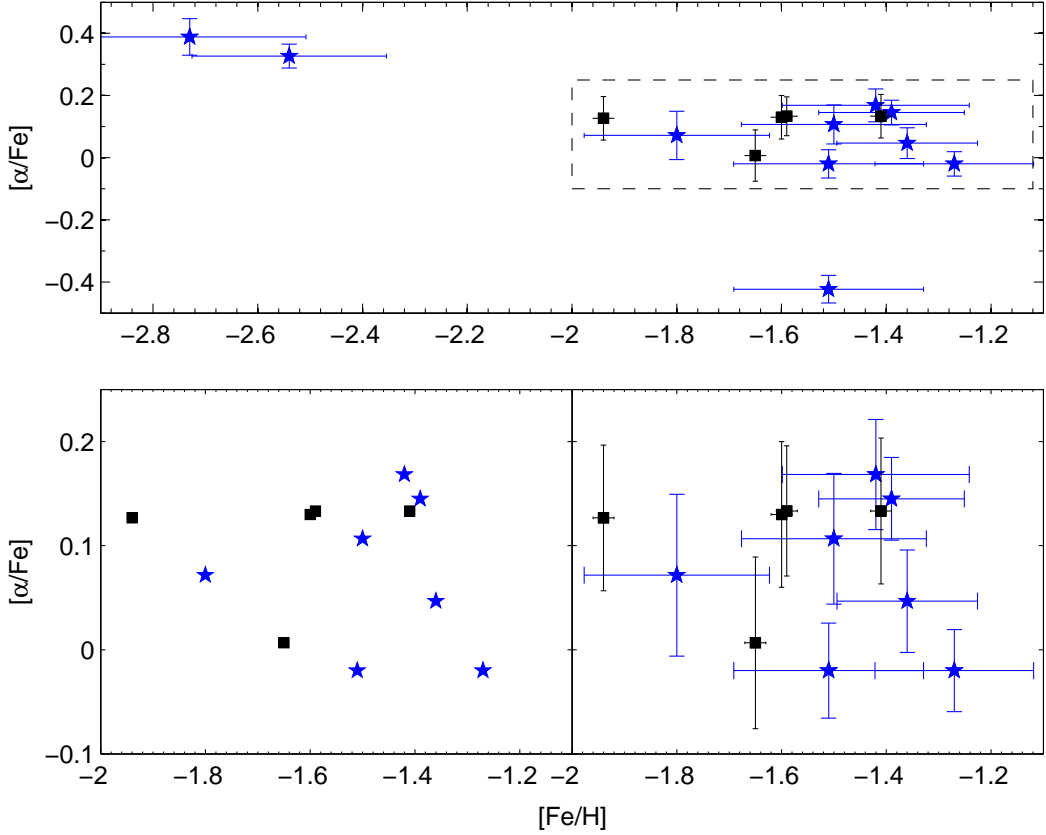}
\end{center}
\caption{Our measured [$\alpha$/Fe] (blue stars) shown together with the data from S03 (black 
squares). The bottom panels are blow-ups of the region enclosed by the dashed quadrangle in the 
top panel. In the bottom left panel errorbars were omitted for clarity. [{\em See the electronic edition 
of the Journal for a color version of this figure.}]}
\end{figure}
\begin{figure}
\begin{center}
\includegraphics[width=14cm]{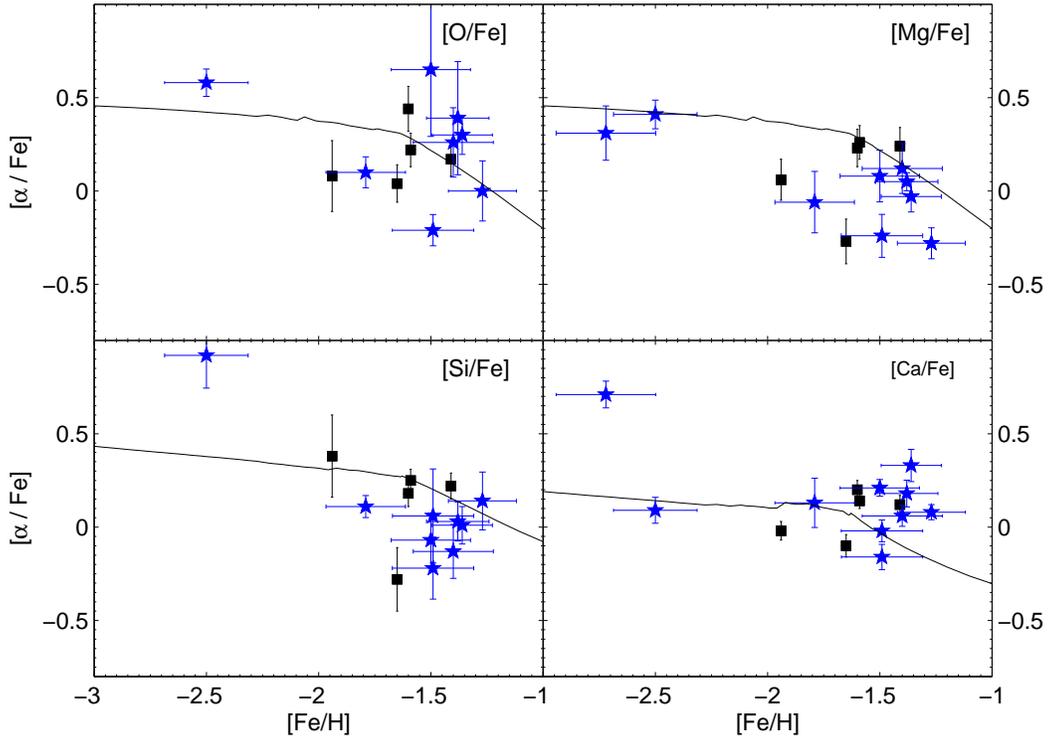}
\end{center}
\caption{Our measured abundance ratios for the 
$\alpha$-elements O, Mg, Si and Ca, 
shown as blue stars, in comparison with the model calculations from LMC06 
(solid lines). Also plotted are the five data points from S03 (black squares). 
[{\em See the electronic edition 
of the Journal for a color version of this figure.}]}
\end{figure}
\begin{figure}
\begin{center}
\includegraphics[width=14cm]{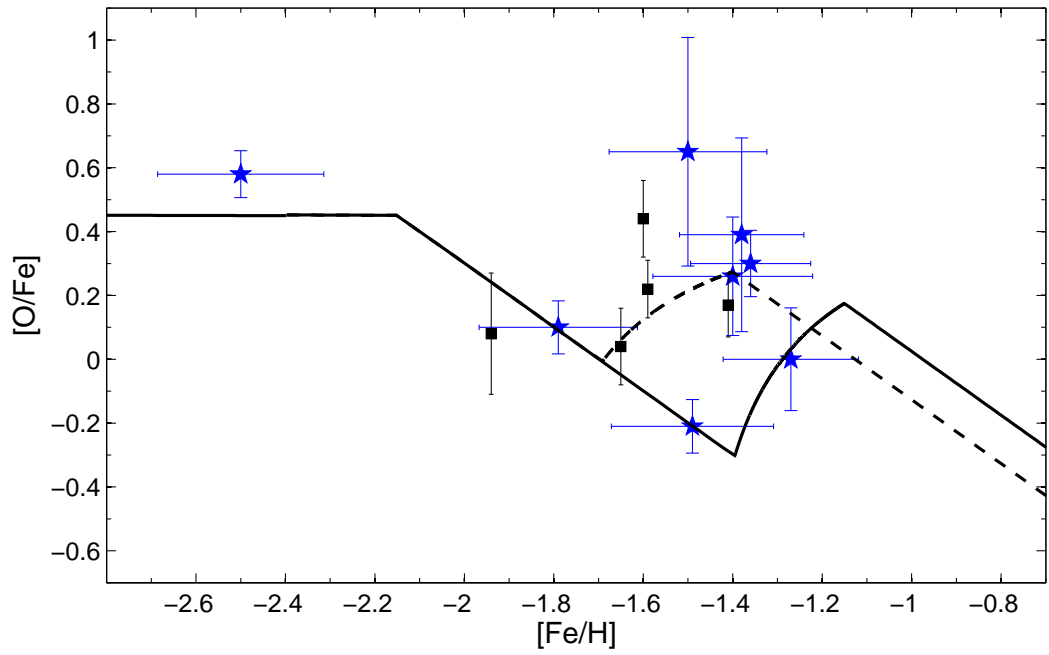}
\end{center}
\caption{Chemical evolution models from Gilmore \& Wyse (1991), which incorporate an initial SF burst, an extended quiescent phase, and a second successive burst. The solid and dashed line 
adopt different onsets of SF. The models were offset 
by $-$0.4 dex in [Fe/H] w.r.t. the original predictions from Gilmore \& Wyse (1991) to match the observed 
element ratios from this work (blue stars) and S03 (black squares). 
[{\em See the electronic edition of the Journal for a color version of this figure.}]}
\end{figure}
\clearpage
\begin{table}
\begin{center}
\caption{Radial velocity members in Carina.}   
\begin{footnotesize}
\begin{tabular}{ccccccccc}
\hline
\hline
Star$^a$ & {$\alpha$ (J2000)} & {$\delta$ (J2000)} & {V} & {B$-$V} & {V$-$K} & 
 {v$_{\rm r}$ [km\,s$^{-1}$]}  & {S/N} \\
\hline
LG04b\_004260 & 06 40 30 & $-$50 59 15 & 17.75 &  0.99 &  3.17 &   215.6 & 18 \\
LG04c\_006477 & 06 41 21 & $-$51 03 43 & 17.71 &  1.31 &  2.71 &  219.5 & 17 \\
LG04a\_000377 & 06 41 59 & $-$50 51 13 & 17.78 &  1.30 &  3.23 &   219.5 & 18 \\
LG04a\_001556 & 06 42 17 & $-$50 55 55 & 17.82 &  1.25 &  3.10 &  223.1 & 19 \\
LG04a\_001826 & 06 40 58 & $-$50 53 35 & 17.76 &  1.29 &  2.88 &  224.8 & 18 \\
LG04c\_000777 & 06 41 04 & $-$51 01 35 & 17.55 &  1.31 &  3.02 & 225.1 & 20 \\
LG04a\_002181 & 06 41 39 & $-$50 49 58 & 17.62 &  1.36 &  2.88 &   230.6 & 21 \\
LG04d\_006628 & 06 41 37 & $-$51 01 43 & 17.95 &  1.28 &  2.90 & 231.1 & 23 \\
LG04c\_000951 & 06 39 55 & $-$50 57 36 & 17.79 &  1.19 &  2.72 &   233.2 & 15 \\
LG04c\_000626 & 06 40 47 & $-$51 06 03 & 17.60 &  1.37 &  3.10 &  237.6 & 16 \\
\hline
\end{tabular}
\end{footnotesize}
\end{center}
{\footnotesize \hspace{3cm} Note. --- The given colors are dereddened values 
in the JC system. \\
\mbox{ }\hspace{3cm} $^a$The nomenclature is such that LG04a--d designates the EIS-fields targeting Carina,  followed\\
\mbox{ }\hspace{3cm}  by the number in the respective input catalog (P.Lynam [EIS-team], 2003, private communication).}
\end{table}
\begin{table}
\begin{center}
\caption{Linelist}
\begin{footnotesize}
\begin{tabular}{cccrrrrrrrrrrrrr}
\hline
\hline
 & {Wavelength} & {EP} & & \multicolumn{11}{c}{Equivalent Width [m\AA]}\\
 \cline{5-15}
 {\raisebox{1.5ex}[-1.5ex]{Ion}}& {[\AA]}& {[eV]}& {\raisebox{1.5ex}[-1.5ex]{$\log\,gf$}} & 
 {Arcturus} &  {4260$^{\,a}$} &  {6477$^{\,a}$} & 
 {377$^{\,a}$} &  {1556$^{\,a}$} &  {1826$^{\,a}$} &  {777$^{\,a}$} &  {2181$^{\,a}$} 
&  {6628$^{\,a}$} &  {951$^{\,a}$} &  {626$^{\,a}$} \\
\hline
Fe\,{\sc i} & 4802.88 &  3.69 & $-$1.64 &  89.8 &  38.8 &  35.0 &  79.3 &  20.5 &  37.5 &  39.6 &  43.5 &  21.5 & \nodata & \nodata \\
Fe\,{\sc i} & 4839.54 &  3.27 & $-$1.90 & 102.4 &  34.0 &  42.1 &  69.8 &  44.6 &  60.2 &  54.3 &  59.5 &  30.5 &    29.3 &    38.1 \\
Fe\,{\sc i} & 4871.32 &  2.85 & $-$0.43 & 246.9 &  91.8 & 102.5 & 120.8 & 126.6 & 114.5 & 125.6 & 141.1 &  82.7 &    74.5 &   116.2 \\
Fe\,{\sc i} & 4903.31 &  2.88 & $-$0.67 & 206.6 &  88.9 & 107.7 & 109.9 & 107.9 &  95.6 &  97.0 & 112.4 &  71.7 &    36.1 &    70.2 \\
Fe\,{\sc i} & 4918.99 &  2.85 & $-$0.41 & 250.5 & 136.1 & 127.2 & 142.9 & 137.5 & 141.2 & 151.0 & 142.4 &  95.0 &    74.2 &    98.4 \\
\hline
\end{tabular}
\end{footnotesize}
\end{center}
{\footnotesize Note. --- This Table is published in its entirety in the electronic edition of the
 {\it Astronomical Journal}.\\
 $^a$For the sake of readability, only the catalogue-numbers are given here, omitting 
the prefixes ``LG04[abcd]\_'' (cf. Table~1)}
\end{table}
\begin{table}
\begin{center}
\caption{Spectroscopic and Photometric model atmosphere parameters}
\begin{footnotesize}
\begin{tabular}{cccccccc}
\hline
\hline
  & {T$_{\mathrm{eff}}$} & {$\sigma$\,T$_{\mathrm{eff}}$} &  {T$_{\mathrm{eff}}$} &  {$\xi$} & 
 {$\log\,g$} &  {[Fe/H]} & {} \\
 {\raisebox{1.5ex}[-1.5ex]{Star}}  & \multicolumn{2}{c}{Photometric} &  {Spectroscopic} & 
 {[km/s]} &  {cgs} &  {(CaT)} &  {\raisebox{1.5ex}[-1.5ex]{[$M$/H]}} \\
\hline
LG04b\_004260 & 4091 & 62 & 4594 & 1.52 & 1.53 & $-$1.63 & $-$1.50 \\
LG04c\_006477 & 4397 & 80 & 4600 & 1.54 & 1.49 & $-$1.55 & $-$1.42 \\
LG04a\_000377 & 4060 & 60 & 4511 & 1.74 & 1.39 & $-$1.45 & $-$1.50 \\
LG04a\_001556 & 4134 & 64 & 4570 & 1.75 & 1.68 & $-$1.41 & $-$1.30 \\
LG04a\_001826 & 4271 & 73 & 4384 & 1.74 & 0.84 & \nodata & $-$1.46 \\
LG04c\_000777 & 4183 & 67 & 4205 & 1.25 & 0.85 & \nodata & $-$1.38 \\
LG04a\_002181 & 4279 & 73 & 4317 & 1.66 & 1.26 & $-$1.39 & $-$1.39 \\
LG04d\_006628 & 4260 & 72 & 4717 & 1.40 & 2.02 & \nodata & $-$1.81 \\
LG04c\_000951 & 4387 & 80 & 4395 & 1.66 & 0.90 & \nodata & $-$2.74 \\
LG04c\_000626 & 4130 & 64 & 4420 & 1.70 & 1.80 & $-$2.06 & $-$2.57 \\
\hline
\end{tabular}
\end{footnotesize}
\end{center}
\end{table}
\begin{table}
\begin{center}
\caption{Error analysis for two representative stars -- uncorrelated contributions}
\begin{footnotesize}
\begin{tabular}{lccccccccc}
\hline
\hline
 & \multicolumn{2}{c}{$\Delta$\,T$_{\mathrm{eff}}$} & \multicolumn{2}{c}{$\Delta$\,log\,$g$} & 
\multicolumn{2}{c}{$\Delta$\,$\xi$} & \multicolumn{2}{c}{$\Delta$\,[$M$/H]} & {} \\
{\raisebox{1.5ex}[-1.5ex]{Ion}} & {+100\,K} &  {$-$100\,K} &  {+0.2} &  {$-$0.2} & 
 {+0.1} &  {$-$0.1} &  {+0.2} &  {$-$0.2} &  {\raisebox{1.5ex}[-1.5ex]{$\alpha$}}  \\
\hline
\multicolumn{10}{c}{}\\
\multicolumn{10}{c}{\raisebox{1.5ex}[-1.5ex]{LG04c\_000777}}\\
\hline
O\,{\sc i}   &$-$0.010  & 0 & +0.090 & $-$0.080 & $-$0.010 & 0 & +0.060 & $-$0.070 & +0.080  \\
Na\,{\sc i}  & +0.090   & $-$0.095 & $-$0.035 &+0.040 & $-$0.010 & +0.015 & $-$0.025 & +0.045 & $-$0.045 \\
Mg\,{\sc i}  & $-$0.039 & +0.028 & $-$0.018 & +0.017 & +0.049 & $-$0.051 & $-$0.012 & +0.011 & $-$0.036   \\
Si\,{\sc i}  & $-$0.132 & +0.138 & +0.032 & $-$0.033 & +0.069 & $-$0.065 & +0.025 & $-$0.023 & +0.010   \\
Ca\,{\sc i}  & +0.023   & $-$0.054 & $-$0.025 & +0.023 & +0.021 & $-$0.021 & $-$0.035 & +0.027 & $-$0.037  \\
Ti\,{\sc i}  & +0.094   & $-$0.125 & $-$0.006 & +0.011 & +0.033 & $-$0.030 & $-$0.034 & +0.029 & $-$0.035  \\
Ti\,{\sc ii} & $-$0.130 & +0.113 & +0.056 & $-$0.059 & +0.019 & $-$0.015 & +0.041 & $-$0.045 & +0.056   \\
Fe\,{\sc i}  & +0.116   & $-$0.081  & $-$0.002 & +0.003 & $-$0.075 & +0.075 & +0.005 & $-$0.001 & +0.040   \\
Fe\,{\sc ii} & $-$0.112 & +0.162 & +0.065 & $-$0.070 & $-$0.053 & +0.058 & +0.070 & $-$0.062 & +0.125   \\
\hline
\multicolumn{10}{c}{}\\
\multicolumn{10}{c}{\raisebox{1.5ex}[-1.5ex]{LG04d\_006628}}\\
\hline
O\,{\sc i} & $-$0.010 & +0.010 & +0.080 & $-$0.090 & 0 & 0 & +0.070 & $-$0.070 & +0.090  \\
Na\,{\sc i} &  \nodata  &  \nodata &  \nodata &  \nodata &  \nodata & \nodata  & \nodata  &  \nodata &  \nodata   \\
Mg\,{\sc i} & $-$0.040 & +0.054 & $-$0.011 & +0.023 & +0.010 & $-$0.003 & +0.006 & $-$0.006 & +0.012  \\
Si\,{\sc i} & $-$0.110 & +0.114 & +0.029 & $-$0.027 & +0.040 & $-$0.043 & +0.016 & $-$0.026 & +0.032 \\
Ca\,{\sc i} & $-$0.044 & +0.039 & $-$0.001 & +0.003 & +0.020 & $-$0.020 & +0.006 & $-$0.009 & +0.001  \\
Ti\,{\sc i} & +0.022 & $-$0.041 & $-$0.005 & $-$0.003 & +0.024 & $-$0.025 & $-$0.001 & +0.005 & $-$0.015   \\
Ti\,{\sc ii} & $-$0.115 & +0.136 & +0.096 & $-$0.082 & 0 & 0 & +0.056 & $-$0.064 & +0.092  \\
Fe\,{\sc i} & +0.140 & $-$0.134 & $-$0.019 & +0.017 & $-$0.040 & +0.043 & $-$0.016 & +0.026 & $-$0.022   \\
Fe\,{\sc ii} & 0 & +0.051 & +0.084 & $-$0.067 & $-$0.023 & +0.021 & +0.034 & $-$0.034 & +0.063 \\
\hline
\end{tabular}
\end{footnotesize}
\end{center}
\end{table}
\begin{table}
\begin{center}
\caption{Abundance results for the analysed Carina red giants}
\begin{footnotesize}
\begin{tabular}{lrcrccrcrccrcrc}
\hline
\hline
 & {} &  {} &  {} &  {} &  {} &  {} & {} &  {}& 
             {} & {} &  {} &  {} &  {} & {} \\
            {} & \multicolumn{4}{c}{LG04a\_002181} &  {} & \multicolumn{4}{c}{LG04b\_004260} &  {} 
                      & \multicolumn{4}{c}{LG04c\_006477} \\
\cline{2-5}\cline{7-10}\cline{12-15}\\
 {\raisebox{1.5ex}[-1.5ex]{Element}} &  {\raisebox{1.5ex}[-1.5ex]{[X/Fe]}} &  {\raisebox{1.5ex}[-1.5ex]{$\sigma_{\rm EW}$}} &  {\raisebox{1.5ex}[-1.5ex]{N}} &  {\raisebox{1.5ex}[-1.5ex]{$\sigma_{\rm tot}$}} &  {} & 
                                               {\raisebox{1.5ex}[-1.5ex]{[X/Fe]}} &  {\raisebox{1.5ex}[-1.5ex]{$\sigma_{\rm EW}$}} &  {\raisebox{1.5ex}[-1.5ex]{N}} &  {\raisebox{1.5ex}[-1.5ex]{$\sigma_{\rm tot}$}} &  {} & 
				               {\raisebox{1.5ex}[-1.5ex]{[X/Fe]}} &  {\raisebox{1.5ex}[-1.5ex]{$\sigma_{\rm EW}$}} &  {\raisebox{1.5ex}[-1.5ex]{N}} &  {\raisebox{1.5ex}[-1.5ex]{$\sigma_{\rm tot}$}}\\
\hline
O\,{\sc i}     &    0.39 &  0.40 &  2 & 0.30 & & $-$0.21 & 0.05 &  1 & 0.08 & &    0.26 & 0.24    &  2 & 0.19 \\
Na\,{\sc i}    & $-$0.31 &  0.39 &  2 & 0.29 & & $-$0.86 & 0.05 &  1 & 0.10 & & $-$0.12 & 0.38    &  3 & 0.24 \\
Mg\,{\sc i}    &    0.05 &  0.03 &  2 & 0.05 & & $-$0.24 &    0.13 &  2 & 0.11 & &    0.12 & 0.18    &  2 & 0.13 \\
Si\,{\sc i}    &    0.03 &  0.10 &  4 & 0.10 & & $-$0.22 &    0.13 &  2 & 0.17 & & $-$0.13 & 0.05 &  1 & 0.14 \\
Ca\,{\sc i}    &    0.18 &  0.04 & 18 & 0.07 & & $-$0.02 &    0.07 & 15 & 0.06 & &    0.06 & 0.08    & 16 & 0.06 \\
Ti\,{\sc i}    & $-$0.12 &  0.04 & 32 & 0.12 & & $-$0.17 &    0.06 & 16 & 0.09 & &    0.13 & 0.07    & 18 & 0.09 \\
Ti\,{\sc ii}   & $-$0.05 &  0.11 &  6 & 0.11 & &    0.00 &    0.07 &  5 & 0.02 & &    0.05 & 0.05    &  5 & 0.10 \\
Fe\,{\sc i}/H  & $-$1.38 &  0.22 & 97 & 0.14 & & $-$1.49 &    0.16 & 96 & 0.18 & & $-$1.40 & 0.16    & 97 & 0.18 \\
Fe\,{\sc ii}/H & $-$1.38 &  0.18 &  7 & 0.27 & & $-$1.48 &    0.13 &  4 & 0.20 & & $-$1.37 & 0.12    &  6 & 0.19 \\
\hline
 & & & & & & & & & & & & & &  \\
 & \multicolumn{4}{c}{LG04a\_000377} & & \multicolumn{4}{c}{LG04a\_001556} & & \multicolumn{4}{c}{LG04a\_001826} \\
\cline{2-5}\cline{7-10}\cline{12-15}\\\
\raisebox{1.5ex}[-1.5ex]{Element} & \raisebox{1.5ex}[-1.5ex]{[X/Fe]} & \raisebox{1.5ex}[-1.5ex]{$\sigma_{\rm EW}$} & \raisebox{1.5ex}[-1.5ex]{N} & \raisebox{1.5ex}[-1.5ex]{$\sigma_{\rm tot}$} & & 
                                    \raisebox{1.5ex}[-1.5ex]{[X/Fe]} & \raisebox{1.5ex}[-1.5ex]{$\sigma_{\rm EW}$} & \raisebox{1.5ex}[-1.5ex]{N} & \raisebox{1.5ex}[-1.5ex]{$\sigma_{\rm tot}$} & & 
				    \raisebox{1.5ex}[-1.5ex]{[X/Fe]} & \raisebox{1.5ex}[-1.5ex]{$\sigma_{\rm EW}$} & \raisebox{1.5ex}[-1.5ex]{N} & \raisebox{1.5ex}[-1.5ex]{$\sigma_{\rm tot}$} \\
\cline{1-15}  
O\,{\sc i}      &    0.65 & 0.50    &  2 & 0.36 & &    0.00 & 0.20 &  2 & 0.16 & & \nodata & \nodata & \nodata &  \nodata \\
Na\,{\sc i}     & $-$0.30 & 0.40    &  2 & 0.29 & & $-$0.55 & 0.21 &  2 & 0.17 & & $-$0.48 & 0.06 &       1 &    0.11  \\
Mg\,{\sc i}     &    0.08 & 0.19    &  2 & 0.14 & & $-$0.28 & 0.10 &  2 & 0.08 & & $-$0.90 & 0.06 &       1 &    0.09  \\
Si\,{\sc i}     &  $-$0.07 & 0.06 &  1 & 0.14 & &    0.14 & 0.17 &  2 & 0.15 & &      0.06 &	0.28 &       2 &    0.25  \\
Ca\,{\sc i}     &    0.21 & 0.04    & 17 & 0.04 & &    0.08 & 0.07 & 16 & 0.04 & &$-$0.16 &	0.05 &      15 &    0.07  \\
Ti\,{\sc i}     &    0.07 & 0.06    & 26 & 0.08 & & $-$0.22 & 0.05 & 28 & 0.20 & & $-$0.83 &	0.08 &      16 &    0.12  \\
Ti\,{\sc ii}    & $-$0.08 & 0.07    &  5 & 0.23 & & $-$0.17 & 0.13 &  6 & 0.11 & & $-$0.63 &	0.11 &       5 &    0.08  \\
Fe\,{\sc i}/H   & $-$1.50 & 0.16    & 91 & 0.18 & & $-$1.27 & 0.14 & 98 & 0.15 & & $-$1.49 &	0.16 &      95 &    0.18  \\
Fe\,{\sc ii}/H  & $-$1.49 & 0.12    &  5 & 0.20 & & $-$1.29 & 0.14 &  7 & 0.23 & & $-$1.43 &	0.14 &       7 &    0.22  \\
\hline
 & & & & & & & & & & & & & & \\
 & \multicolumn{4}{c}{LG04c\_000777} & & \multicolumn{4}{c}{LG04d\_006628} & & \multicolumn{4}{c}{LG04c\_000951} \\
\cline{2-5}\cline{7-10}\cline{12-15}\\\
\raisebox{1.5ex}[-1.5ex]{Element} &  
                                    \raisebox{1.5ex}[-1.5ex]{[X/Fe]} & \raisebox{1.5ex}[-1.5ex]{$\sigma_{\rm EW}$} & \raisebox{1.5ex}[-1.5ex]{N} & \raisebox{1.5ex}[-1.5ex]{$\sigma_{\rm tot}$}  & & 
				    \raisebox{1.5ex}[-1.5ex]{[X/Fe]} & \raisebox{1.5ex}[-1.5ex]{$\sigma_{\rm EW}$} & \raisebox{1.5ex}[-1.5ex]{N} & \raisebox{1.5ex}[-1.5ex]{$\sigma_{\rm tot}$}  & & 
				    \raisebox{1.5ex}[-1.5ex]{[X/Fe]} & \raisebox{1.5ex}[-1.5ex]{$\sigma_{\rm EW}$} & \raisebox{1.5ex}[-1.5ex]{N} & \raisebox{1.5ex}[-1.5ex]{$\sigma_{\rm tot}$} \\  
\cline{1-15}
O\,{\sc i}     &    0.30 & 0.08 &  1 & 0.10 & &    0.10 & 0.06 &       1 &    0.08 & & \nodata & \nodata & \nodata & \nodata  \\
Na\,{\sc i}    & $-$0.32 &    0.13 &  2 & 0.12 & & \nodata & \nodata & \nodata & \nodata & & \nodata & \nodata & \nodata & \nodata  \\
Mg\,{\sc i}    & $-$0.03 &    0.20 &  2 & 0.08 & & $-$0.06 &0.13 &       1 &    0.16 & &	0.31 & 0.12 &       1 &    0.15  \\
Si\,{\sc i}    &    0.01 &    0.24 &  3 & 0.10 & &    0.11 & 0.04 &       1 &    0.06 & & \nodata & \nodata & \nodata & \nodata  \\
Ca\,{\sc i}    &    0.33 &    0.06 & 17 & 0.09 & &    0.13 &    0.10 &      11 &    0.13 & &	0.71 &    0.05 &       9 &    0.07  \\
Ti\,{\sc i}    & $-$0.09 &    0.06 & 29 & 0.08 & & $-$0.25 &    0.08 &      11 &    0.14 & &	0.10 &    0.14 &       8 &    0.09  \\
Ti\,{\sc ii}   &    0.17 &    0.15 &  5 & 0.15 & &    0.18 &    0.17 &       4 &    0.14 & &	0.04 &    0.12 &       3 &    0.10  \\
Fe\,{\sc i}/H  & $-$1.36 &    0.12 & 90 & 0.13 & & $-$1.79 &    0.15 &      86 &    0.18 & & $-$2.72 &    0.16 &      67 &    0.22  \\
Fe\,{\sc ii}/H & $-$1.35 &    0.19 &  4 & 0.29 & & $-$1.79 &    0.12 &       7 &    0.14 & & $-$2.72 &    0.12 &       4 &    0.13  \\
\hline
 & & & & & & & & & & & & & & \\
 & \multicolumn{4}{c}{LG04c\_00626} & & & & & & & & & & \\
\cline{2-5}\\
\raisebox{1.5ex}[-1.5ex]{Element} &  
                                    \raisebox{1.5ex}[-1.5ex]{[X/Fe]} & \raisebox{1.5ex}[-1.5ex]{$\sigma_{\rm EW}$} & \raisebox{1.5ex}[-1.5ex]{N} & \raisebox{1.5ex}[-1.5ex]{$\sigma_{\rm tot}$}   
				  & & & & & & & & & & \\
\cline{1-5}
O\,{\sc i}     &    0.58 & 0.05 &  1 &    0.07 & & & & & & & & & &  \\
Na\,{\sc i}    &    0.02 & 0.05 &  1 & 0.10 & & & & & & & & & &  \\
Mg\,{\sc i}    &    0.41 &    0.09 &  2 &    0.08 & & & & & & & & & &  \\
Si\,{\sc i}    &    0.92 &    0.13 &  2 &    0.18 & & & & & & & & & &  \\
Ca\,{\sc i}    &    0.09 &    0.11 &  9 &    0.07 & & & & & & & & & &  \\
Ti\,{\sc i}    &    0.23 &    0.05 &  9 &    0.10 & & & & & & & & & &  \\
Ti\,{\sc ii}   &    0.40 &    0.08 &  3 &    0.03 & & & & & & & & & &  \\
Fe\,{\sc i}/H  & $-$2.50 &    0.15 & 79 &    0.19 & & & & & & & & & &  \\
Fe\,{\sc ii}/H & $-$2.54 &    0.11 &  4 &    0.16 & & & & & & & & & &  \\
\cline{1-5}
\end{tabular}
\end{footnotesize}
\end{center}
\end{table}

\begin{thebibliography}{}
\bibitem[Akerman et al.(2004)]{2004A&A...414..931A} Akerman, C.~J., Carigi, 
L., Nissen, P.~E., Pettini, M., \& Asplund, M.\ 2004, \aap, 414, 931 
%
\bibitem[Alonso et al. 2001]{Alonso2001} Alonso, A., Arribas, S., \& Mart\'inez-Roger, C. 2001, 
	AJ, 376, 1039
%
\bibitem[Anders & Grevesse 1989]{Anders1989} Anders, E., \& Grevesse, N. 1989, Geochim. Cosmochim. Acta, 53, 197
%
\bibitem[Argast et al. 2000]{Argast2000} Argast, D., Samland, M., Gerhard, O.~E., \& Thielemann, F.-K. 2000, A\&A, 356, 873
%
\bibitem[Argast et al. 2002]{Argast2002} Argast, D., Samland, M., Thielemann, F.-K., \& Gerhard, O.~E. 2002, A\&A, 388, 842
%
\bibitem[Armandroff & Da Costa 1991]{Armandroff1991} Armandroff, T.E., \& Da Costa, G.S. 1991, AJ, 101, 1329
%
\bibitem[Asplund et al. 2005]{Asplund2005} Asplund, M., Grevesse, N., \& Sauval, A.~J. 2005, in ASP Conf. Ser.,  	
	Cosmic Abundances as Records of Stellar Evolution and Nucleosynthesis, 
	ed. T.G. Barnes III \& F.N. Bash (San Francisco: ASP), 236,  25
%
\bibitem[Bai et al.(2004)]{2004A&A...425..671B} Bai, G.~S., Zhao, G., Chen, 
Y.~Q., Shi, J.~R., Klochkova, V.~G., Panchuk, V.~E., Qiu, H.~M., \& Zhang, 
H.~W.\ 2004, \aap, 425, 671 
%
\bibitem[Bard & Kock 1994]{Bard1994} Bard, A., \& Kock, M. 1994, A\&A, 282, 1014
%
\bibitem[Battaglia et al.(2008)]{2008arXiv0710.0798B} Battaglia, G., Irwin, 
M., Tolstoy, E., Hill, V., Helmi, A., Letarte, B., \& Jablonka, P.\ 2008, MNRAS, 383, 183
%
\bibitem[Bessel & Brett 1988]{Bessel1988} Bessel, M.~S.\& Brett, J.~M. 1988, PASP, 100, 1134
%
\bibitem[Blackwell et al. 1995]{Blackwell1995}Blackwell, D.~E., Smith, G., \& Lynas-Gray, A.~E. 1995, A\&A, 303, 575
%
\bibitem[Boesgaard et al.(1999)]{1999AJ....117..492B} Boesgaard, A.~M., 
King, J.~R., Deliyannis, C.~P., \& Vogt, S.~S.\ 1999, \aj, 117, 492 
%
\bibitem[Bosler et al.(2007)]{2007MNRAS.378..318B} Bosler, T.~L., 
Smecker-Hane, T.~A., \& Stetson, P.~B.\ 2007, \mnras, 378, 318 
%
\bibitem[Cardelli et al. 1989]{Cardelli1989}  Cardelli, J.~A., Clayton, G.~C., \& Mathis, J.~S. 1989, ApJ, 345, 245
%
\bibitem[Carney 1996]{Carney1996} Carney, B. 1996, PASP, 108, 90
%
\bibitem[Carney et al. 1997]{Carney1997} Carney, B.~W., Wright, J.~S., Sneden, C., Laird, J.~B., Aguilar, 
L.~A., \& Latham, D.~W. 1997, AJ, 114, 363
%
\bibitem[Carollo et al.(2007)]{2007Natur.450.1020C} Carollo, D., et al.\ 
2007, \nat, 450, 1020
%
\bibitem[Carrera et al.(2007)]{2007AJ....134.1298C} Carrera, R., Gallart, 
C., Pancino, E., \& Zinn, R.\ 2007, \aj, 134, 1298 
%
\bibitem[Carretta & Gratton 1997]{Carretta1997} Carretta, E., \& Gratton, R. 1997, A\&AS, 121, 95
%
\bibitem[Carretta 2006]{Carretta2006} Carretta, E. 2006, AJ, 131, 1766
%
\bibitem[Castelli2003]{Castelli2003} Castelli, F., \& Kurucz, R.~L. 2003, in IAU Symp. 210, Modelling of Stellar Atmospheres, 
eds. N.E. Piskunov, W.W. Weiss, \& D.F. Gray (San Francisco: ASP), A20 (astro-ph/0405087)
%
 \bibitem[Cayrel et al.(2004)]{2004A&A...416.1117C} Cayrel, R., et al.\ 
2004, \aap, 416, 1117 
%
\bibitem[Cohen \& Mel{\'e}ndez(2005)]{2005AJ....129..303C} Cohen, J.~G., \& 
Mel{\'e}ndez, J.\ 2005, \aj, 129, 303 
%
\bibitem[Cohen et al.(2007)]{2007ApJ...659L.161C} Cohen, J.~G., McWilliam, 
A., Christlieb, N., Shectman, S., Thompson, I., Melendez, J., Wisotzki, L., 
\& Reimers, D.\ 2007, \apjl, 659, L161 
%
\bibitem[Cutri 2003]{Cutri 2003} Cutri R.~M. 2003, Explanatory Supplement to the 2MASS All-Sky Data Release, 
\url{http://www.ipac/caltech.edu/2mass/releases/allsky/doc/explsup.html}
%
\bibitem[Decressin et al.(2007)]{Decressin2007} Decressin, T., 
Meynet, G., Charbonnel, C., Prantzos, N., \& Ekstr{\"o}m, S.\ 2007,
\aap, 464, 1029 
%
\bibitem[Fran{\c c}ois et al.(2004)]{2004A&A...421..613F} Fran{\c c}ois, 
P., Matteucci, F., Cayrel, R., Spite, M., Spite, F., \& Chiappini, C.\ 
2004, \aap, 421, 613 
%
\bibitem[Friel et al. 2003]{Friel2003} Friel, E.~D., Jacobson, H.~R., Barrett, E., Fullton, L., Balachandran, S.~C., 
\& Pilachowski, C.~A. 2003, AJ, 126, 2372 
%
\bibitem[Fulbright(2002)]{Fulbright2002} Fulbright, J.~P.\ 2002, \aj, 
123, 404 
%
\bibitem[Fulbright et al. 2006]{Fulbright2006} Fulbright, J.~P., McWilliam, A., \& Rich, R~M. 2006, 
ApJ, 636, 821
%
\bibitem[Fulbright et al.(2007)]{2007ApJ...661.1152F} Fulbright, J.~P., 
McWilliam, A., \& Rich, R.~M.\ 2007, \apj, 661, 1152 
%
\bibitem[Gallagher \& Wyse(1994)]{Gallagher1994} Gallagher, J.S., III,
\& Wyse, R.F.G. 1994, \pasp, 106, 1225
%
\bibitem[Geisler et al. 2005]{Geisler2005} Geisler, D., Smith, V.~V., Wallerstein, G., Gonzalez, G., \& Charbonnel, C. 2005, AJ, 129, 1428
%
\bibitem[Gilmore & Wyse 1998]{Gilmore1998} Gilmore, G., \& Wyse, R.~F.~G. 1998, AJ, 116, 748
%
\bibitem[Gilmore & Wyse 1991]{Gilmore1991} Gilmore, G. \& Wyse, R.F.G. 1991, ApJ, 367, L55 
%
\bibitem[Gilmore  et al.(2007)]{Gimore2007} Gilmore, G., Wilkinson,
M.I., Wyse, R.F.G., Kleyna, J.T., Koch, A., Evans, N.W., \& Grebel,
E.K. 2007, ApJ, 663, 948
%
\bibitem[Gonzalez \& Wallerstein 1999]{Gonzalez99} Gonzalez, G., \& Wallerstein, G. 1999, \aj, 
117, 2286
%
\bibitem[Gratton et al. 2004]{Gratton2004} Gratton,  R., Sneden, C., \& Carretta, E. 2004, ARA\&A, 42, 385
%
\bibitem[Grebel \& Richtler (1992)]{Grebel1992} Grebel, ,E.~K., \& Richtler, T., 1992, \aap, 253, 359
%
\bibitem[Grebel \& Gallagher(2004)]{Grebel2004} Grebel, E.~K., \& 
Gallagher, J.~S., III 2004, \apjl, 610, L89 
%
\bibitem[Grebel et al.(2003)]{Grebel03} Grebel, E.~K., Gallagher,
J.~S., III, \& Harbeck, D.\ 2003, \aj, 125, 1926
%
\bibitem[Grevesse & Sauval]{Grevesse1999} Grevesse, N., \& Sauval, A.~J. 1999, A\&A, 347, 348
%
\bibitem[Harbeck et al.(2001)]{Harbeck01} Harbeck, D., et al.\
2001, \aj, 122, 3092
%
\bibitem[Helmi et al.(2006)]{2006ApJ...651L.121H} Helmi, A., et al.\ 2006, 
\apjl, 651, L121 
%
\bibitem[Hinkle et al. 2000]{Hinkle2000} Hinkle, K., Wallace, L., Valenti, J., \& Harmer, D. 2000, 
	Visible and Near Infrared Atlas of the Arcturus Spectrum 3727-9300\AA\ 
	(San Francisco: ASP), online atlas at 
	\url{ftp://ftp.noao.edu/catalogs/arcturusatlas/visual/}
%
\bibitem[Israelian et al.(1998)]{1998ApJ...507..805I} Israelian, G., 
Garc{\'{\i}}a L{\'o}pez, R.~J., \& Rebolo, R.\ 1998, \apj, 507, 805 
%
\bibitem[Ivans et al.(2003)]{2003ApJ...592..906I} Ivans, I.~I., Sneden, C., 
James, C.~R., Preston, G.~W., Fulbright, J.~P., H{\"o}flich, P.~A., Carney, 
B.~W., \& Wheeler, J.~C.\ 2003, \apj, 592, 906 
%
\bibitem[Johnson 2002]{Johnson2002} Johnson, J.~A. 2002, ApJSS, 139, 219
%
\bibitem[Johnson et al. 2006]{Johnson2006} Johnson, J.~A., Ivans, I.~I., \& Stetson, P.~B. 2006, ApJ, 640, 801
%
\bibitem[kleyna2003]{k7} Kleyna, J.~T., Wilkinson, M.~I., Gilmore, G., \& Evans, N.~W. 2003, 
ApJ, 588, L21
%
\bibitem[Kniazev et al.(2005)]{Kniazev2005} Kniazev, A.~Y., Grebel, 
E.~K., Pustilnik, S.~A., Pramskij, A.~G., \& Zucker, D.~B.\ 2005, \aj,
130, 1558 
%
\bibitem{koch06} Koch, A., Grebel, E.~K., Wyse, R.~F.~G., Kleyna, J.~T., Wilkinson, M.~I., 
Harbeck, D.~R., Gilmore, G.~F., \& Evans, N.~W. 2006, \aj, 131, 895 (Paper\,I)
%
\bibitem[Koch et al.(2007)]{Koch07} Koch, A., Grebel, E.~K.,
Kleyna, J.~T., Wilkinson, M.~I., Harbeck, D.~R., Wyse, R.~F.~G., \& Evans, N.~W., 
2007, AJ, 133, 270
%
\bibitem[Koch&McW]{KMcW07} Koch, A., \& McWilliam, A. 2008, AJ, submitted
%
\bibitem[Kraft & Ivans 2003]{Kraft2003}  Kraft, R.~P., \& Ivans, I.I. 2003, PASP,
	115, 143
%
\bibitem[Langer et al. 1993]{Langer1993} Langer, G.~E.,  Hoffmann R.~D., Sneden, C. 1993, PASP, 105, 301
%
\bibitem[Lanfranchi & Matteucci 2004]{Lanfranchi2004} Lanfranchi, 
G.~A., \& Matteucci, F.\ 2004, MNRAS, 351, 1338
%
\bibitem[Lanfranchi et al. 2006a]{Lanfranchi2006} Lanfranchi, G.~A., Matteucci, F., \& Cescutti, G.  2006a, MNRAS, 365, 477
%
\bibitem[Lanfranchi et al. 2006b]{LMC06} Lanfranchi, G.~A.,  Matteucci, F., \& Cescutti, G. \ 2006b, A\&A, 453, L67 (LMC06)
%
\bibitem[Langer & Hoffman 1995]{Langer1995} Langer, G.~E., \& Hoffmann R.~D. 1995, PASP, 107, 1177
%
\bibitem[Letarte et al. 2006]{Letarte2006} Letarte, B., Hill, V., Jablonka, P., Tolstoy, E., \& Meylan, G. 2006, 
A\&A, 453, 547
%
\bibitem[Luck & Bond 1985]{Luck1985}  Luck, R.~E., \& Bond, H.~E. 1985, ApJ, 292, L559
%
\bibitem[Mateo 1998]{Mateo1998} Mateo, M. 1998, ARA\&A, 36, 435
%
\bibitem[Majewski et al. 2000]{Majewski2000} Majewski, S.~R., 
Ostheimer, J.~C., Patterson, R.~J., Kunkel, W.~E., Johnston, K.~V., \& 
Geisler, D.\ 2000, AJ, 119, 760
%
\bibitem[Matteucci \& Greggio 1986]{mg86} Matteucci, F., \& Greggio, L. 1986,  A\&A 154, 279
%
\bibitem[Matteucci 2003]{matteucci2003} Matteucci, F. 2003, Ap\&SS, 284, 539
%
\bibitem[McWilliam et al. 1995]{McWilliam1995} McWilliam, A., Preston, G.~W., Sneden, C. \& Searle, L. 1995,  AJ, 109, 275
%
\bibitem[McWilliam 1997]{McWilliam1997} McWilliam, A. 1997, ARA\&A, 35, 503
%
\bibitem[Mighell 1997]{Mighell1997}  Mighell, K.J. 1997, AJ, 114, 1458
%
\bibitem[Miller \& Scalo(1979)]{1979ApJS...41..513M} Miller, G.~E., \& 
Scalo, J.~M.\ 1979, \apjs, 41, 513 
%
\bibitem[Mishenina et al.(2000)]{2000A&A...353..978M} Mishenina, T.~V., 
Korotin, S.~A., Klochkova, V.~G., \& Panchuk, V.~E.\ 2000, \aap, 353, 978 
%
\bibitem[Monaco et al. 2005]{mon05} Monaco, L., Bellazzini, M., Bonifacio, P., Ferraro, F.~R., 
Marconi, G., Pancino, E., Sbordone, L., \& Zaggia, S. 2005, \aap, 441, 141
%
\bibitem[Monelli et al. 2003]{Monelli2003} Monelli, M., et al.\ 2003, AJ, 126, 218
%
\bibitem[Nissen \& Schuster(1997)]{1997A&A...326..751N} Nissen, P.~E., \& 
Schuster, W.~J.\ 1997, \aap, 326, 751 
%
\bibitem[Nonino et al. 1999]{Nonino1999} Nonino, M., et al.\ 1999, A\&AS, 137, 51
%
\bibitem[Norris et al.(2002)]{2002ApJ...569L.107N} Norris, J.~E., Ryan, 
S.~G., Beers, T.~C., Aoki, W., \& Ando, H.\ 2002, \apjl, 569, L107 
%
\bibitem[O'Brian et al.  1997]{Obrian1997} O'Brian, T.~R., Wickliff, M.~E., Lawler, J.~E., Whaling, W., \& Brault, J.~W. 1991, 
J. Opt. Soc. Am. B., 8, 1185
%
\bibitem[Peterson et al. 1993]{Peterson1993} Peterson, R.~C, Dalle Ore, C.~M., \& Kurucz, R.~L. 1993
	ApJ, 404, 333
%
\bibitem[Pont et al. 2004]{Pont2004} Pont, F., Zinn, R., 
Gallart, C., Hardy, E., \& Winnick, R.\ 2004, AJ, 127, 840
%
\bibitem[Pritzl et al. (2005)]{Pritzl2005} Pritzl, B.~J., Venn, K.~A., \& Irwin, M.~I. 2005, AJ, 130, 2140
%
\bibitem[Prochaska et al. 2000]{Prochaska2000} Prochaska, J.~X., Naumov, S.~O., Carney, B.~W., McWilliam, A., 
	\& Wolfe, A.~M. 2000, AJ, 120, 2153
%
\bibitem[Rizzi et al. 2003]{Rizzi2003} Rizzi, L., Held, E.~V., 
Bertelli, G., \& Saviane, I.\ 2003, ApJ, 589, L85
%
\bibitem[Rutledge et al. 1997a]{Rutledge1997a} Rutledge, G.A.,  Hesser, J.E., Stetson, P.B., Mateo, M., Simard, L., Bolte, M., Friel, E.D., \& Copin, Y. 1997a, PASP, 109, 883 
%
\bibitem[Rutledge et al. 1997b]{Rutledge1997b} Rutledge, G.A.,  Hesser, J.E., Stetson, P.B. 1997b, PASP, 109, 907
%
\bibitem[Sadakane et al. 2004]{Sadakane2003} Sadakane, K., Arimoto, N., Ikuta, C., Aoki, W., Jablonka, 
P., \& Tajitsu, A. 2004, PASJ, 56, 1041
%
\bibitem[Salaris et al.(1993)]{1993ApJ...414..580S} Salaris, M., Chieffi, 
A., \& Straniero, O.\ 1993, \apj, 414, 580 
%
\bibitem[Schlegel et al. 1998]{Schlegel1998} Schlegel, D.J., Finkbeiner, D.P., \& Davis, M. 1998, ApJ, 500, 525
%
\bibitem[Shetrone et al. 2001]{Shetrone2001} Shetrone, M.~D., C\^ot\'e, 
P., \& Sargent, W.~L.~W.\ 2001, ApJ, 548, 592
%
\bibitem[Shetrone et al. 2003]{Shetrone2003} Shetrone, M.~D., Venn, K.~A., Tolstoy, E., Primas, F., 
Hill, V., \& Kaufer, A. 2003, AJ, 125, 684 (S03)
%
\bibitem[Sivarani et al. 2004]{Sivarani2004} Sivarani, T., et al. 2004, A\&A, 413, 1073
%
\bibitem[Smecker-Hane et al. 1994]{SmeckerHane1994} Smecker-Hane, T.A., Stetson, P.B., Hesser, J.E., \& Lehnert, M.D. 1994, AJ, 108, 507
%
\bibitem[Smecker-Hane et al. 1996]{SmeckerHane1996} Smecker-Hane, 
T.~A., Stetson, P.~B., Hesser, J.~E., \& Vandenberg, D.~A.\ 1996, in
From Stars to Galaxies: the Impact of Stellar Physics on Galaxy Evolution, 
ASP Conf.\ Ser.\ Vol\ 98, eds.\ C.\ Leitherer, U.\ Fritze-von Alvensleben,
\& J.\ Huchra (San Francisco: ASP), 328
%
\bibitem[Smecker-Hane et al. 1999]{SmeckerHane1999}  Smecker-Hane, T.A.,
Mandushev, G.I., Hesser, J.E., Stetson, P.B., Da Costa, G.S, \& 
Hatzidimitriou, D.  1999, in Spectrophotometric Dating of Stars and
Galaxies, ASP Conf.\ Ser.\ Vol.\ 192, eds.\ I.\ Hubeny, S.\ Heap, \& R.\
Cornett (San Francisco: ASP), 159 
%
\bibitem[Sneden 1973]{Sneden1973} Sneden, C. 1973, ApJ, 184, 839
%
\bibitem[Thielemann et al. 1996]{Thielemann1996} Thielemann, F.-K., Nomoto,  K., \& Hashimoto, M. 1996, ApJ, 460 408
%
\bibitem[Thielemann et al. 2001]{Thielemann2001} Thielemann, F.-K., et al. 2001, PrPNP, 46, 5
%
\bibitem[Tinsley 1976]{tinsley76} Tinsley, B. 1976, ApJ, 208, 797
%
\bibitem[Timmes et al. 1995]{Timmes1995} Timmes, F.~X., Woosley, S.~E., \& Weaver, T.~A. 1995, ApJS, 98, 617
%
\bibitem[Tolstoy et al. 2001]{Tolstoy2001} Tolstoy, E., Irwin, M.J., Cole, A.A., Pasquini, L., Gilmozzi, R., \& Gallagher, J.S. 2001, MNRAS, 327, 918
%
\bibitem[Tolstoy et al. 2003]{Tolstoy2003} Tolstoy, E., Venn, K.~A., Shetrone, M.~D., Primas, F.,
 Hill, V., Kaufer, A., \& Szeifert, T. 2003, AJ, 125, 707 (T03)
%
\bibitem[Unavane et al.(1996)]{1996MNRAS.278..727U} Unavane, M., Wyse, 
R.~F.~G., \& Gilmore, G.\ 1996, \mnras, 278, 727 
%
\bibitem[Venn et al. 2004]{Venn2004} Venn, K.~A., Irwin, M.~I., Shetrone, M.~D., Tout, C.~A., Hill, V., 
\& Tolstoy, E. 2004, AJ, 128, 1177
%
\bibitem[Wilkinson et al.(2006)]{Wilkinson06} Wilkinson, M.~I.,
Kleyna, J.~T.,  Gilmore, G.~F., Evans, N.~W., Koch, A., Grebel, E.~K.,
Wyse, R.~F.~G., \& Harbeck, D.\ 2006, The ESO Messenger, 124, 25
%
\bibitem[Woosley 1986]{Woosley1986} Woosley, S.~E. 1986, in Nucleosynthesis and Chemical Evolution, 16th Advanced Course Swiss Society of Astrophysics and Astronomy, ed. 
B. Hauck, A. Maeder \& G. Meynet (Geneva: Geneva Observatory), 74
%
\bibitem[Woosley & Weaver 1995]{Woosley1995} Woosley, S.~E., \& Weaver, T.~A. 1995, ApJS, 101, 181
%
\bibitem[Wyse \& Gilmore 1988]{Wyse1988} Wyse, R.~F.~G., \& Gilmore, G.~F. 1988, AJ, 95, 1404
%
\bibitem[Wyse et al. 2002]{Wyse2002}  Wyse, R.~F.~G.,  2002, New Astron., 7, 395
%
\bibitem[Wyse et al. 2006]{Wyse2006} Wyse, R.~F.~G., Gilmore, G., 
Norris, J.~E., Wilkinson, M.~I., Kleyna, J.~T., Koch, A., Evans, N.~W., 
\& Grebel, E.~K.\ 2006, ApJ, 639, L13 
%
\bibitem[Zhang \& Zhao(2005)]{2005MNRAS.364..712Z} Zhang, H.~W., \& Zhao, 
G.\ 2005, \mnras, 364, 712
%
 \bibitem[Zinn & West 1984]{Zinn1984}  Zinn, R., \& West, M.~J. 1984, ApJS, 55, 45 
%
\end{thebibliography}
\end{document}